\documentclass[fleqn,usenatbib]{mnras}

\usepackage{newtxtext,newtxmath}

\usepackage[T1]{fontenc}
\usepackage{bm}
\usepackage{cleveref}
\crefname{figure}{Fig.}{Figs.}
\crefname{table}{Table}{Tables}

\DeclareRobustCommand{\VAN}[3]{#2}
\let\VANthebibliography\thebibliography
\def\thebibliography{\DeclareRobustCommand{\VAN}[3]{##3}\VANthebibliography}
\newcommand {\ts}{\textsubscript}

\usepackage{graphicx}	
\usepackage{amsmath}	
\usepackage{xspace}
\makeatletter 
  \patchcmd{\NAT@citex}
    {\@citea\NAT@hyper@{%
      \NAT@nmfmt{\NAT@nm}%
      \hyper@natlinkbreak{\NAT@aysep\NAT@spacechar}{\@citeb\@extra@b@citeb}%
      \NAT@date}}
    {\@citea\NAT@nmfmt{\NAT@nm}%
    \NAT@aysep\NAT@spacechar\NAT@hyper@{\NAT@date}}{}{}

  \patchcmd{\NAT@citex}
    {\@citea\NAT@hyper@{%
      \NAT@nmfmt{\NAT@nm}%
      \hyper@natlinkbreak{\NAT@spacechar\NAT@@open\if*#1*\else#1\NAT@spacechar\fi}%
        {\@citeb\@extra@b@citeb}%
      \NAT@date}}
    {\@citea\NAT@nmfmt{\NAT@nm}%
    \NAT@spacechar\NAT@@open\if*#1*\else#1\NAT@spacechar\fi\NAT@hyper@{\NAT@date}}
    {}{}
\makeatother

\newcommand\HI{\ion{H}{I}\xspace} 
\newcommand\HII{\ion{H}{II}\xspace} 
\newcommand\HeI{\ion{He}{I}\xspace} 
\newcommand\HeII{\ion{He}{II}\xspace} 
\newcommand\HeIII{\ion{He}{III}\xspace} 

\title[Accelerating {\sc arepo-rt}]{Adapting {\sc arepo-rt} for Exascale Computing: GPU Acceleration and Efficient Communication}

\author[Zier et al.]{
Oliver Zier,$^{1}$\thanks{E-mail: \href{mailto:ozier@mit.edu}{ozier@mit.edu}}
Rahul Kannan,$^{2}$
Aaron Smith,$^{3}$
Mark Vogelsberger$^{1}$
and Erkin Verbeek$^{4}$
\\
$^{1}$Department of Physics, Kavli Institute for Astrophysics and Space Research, Massachusetts Institute of Technology, Cambridge, MA 02139, USA\\
$^{2}$Department of Physics and Astronomy, York University, 4700 Keele Street, Toronto, ON M3J 1P3, Canada\\
$^{3}$Department of Physics, The University of Texas at Dallas, Richardson, TX 75080, USA\\
$^{4}$Center for Computational Science and Engineering, Massachusetts Institute of Technology, Cambridge, MA 02139, USA
}

\date{Accepted XXX. Received YYY; in original form ZZZ}

\pubyear{2024}

\begin{document}
\label{firstpage}
\pagerange{\pageref{firstpage}--\pageref{lastpage}}
\maketitle

\begin{abstract}
Radiative transfer (RT) is a crucial ingredient for self-consistent modelling of numerous astrophysical phenomena across cosmic history.
However, on-the-fly integration into radiation-hydrodynamics (RHD) simulations is computationally demanding, particularly due to the stringent time-stepping conditions and increased dimensionality inherent in multi-frequency collisionless Boltzmann physics. The emergence of exascale supercomputers, equipped with extensive CPU cores and GPU accelerators, offers new opportunities for enhancing RHD simulations.
We present a novel optimization of {\small AREPO-RT} explicitly tailored for
such high-performance computing environments.
We implement a novel node-to-node communication strategy that utilizes shared memory to substitute intra-node communication with direct memory access.
Furthermore, combining multiple inter-node messages into a single message substantially enhances network bandwidth utilization and performance for large-scale simulations on modern supercomputers.
The single-message node-to-node approach also improves performance on smaller-scale machines with less optimized networks.
Furthermore, by transitioning all RT-related calculations to GPUs, we achieve a significant computational speedup of around 15 for standard benchmarks compared to the original CPU implementation.
As a case study, we perform cosmological RHD simulations of the Epoch of Reionization, employing a similar setup as the {\small THESAN} project. In this context, RT becomes sub-dominant such that even without modifying the core {\small AREPO} codebase, there is an overall threefold improvement in efficiency.
The advancements presented here have broad implications, potentially transforming the complexity and scalability of future simulations for a wide variety of astrophysical studies.
Our work serves as a blueprint for porting similar simulation codes based on unstructured resolution elements to GPU-centric architectures.
\end{abstract}




\section{Introduction}
Since the birth of the Universe, electromagnetic radiation consisting of photons has permeated its expanse.
Shortly after the Big Bang, the Universe was filled with a hot, dense plasma of matter and radiation. 
The constant interactions between electrons and photons through Thompson scattering tightly coupled radiation and baryonic matter, preventing the growth of initial density perturbations within the baryonic component. Consequently, the initial formation of the deepest gravitational potential wells was reliant on dark matter.
After around 380,000 years (redshift $z\approx 1100$), the Universe had expanded enough to cool to approximately $3000$\,K, allowing protons and electrons to recombine to form neutral atomic hydrogen. After this the Universe became transparent to the primordial radiation, which is now observed as the cosmic microwave background \citep[CMB;][]{alpher1948relative,penzias1979measurement} with a present-day temperature of approximately $2.725$\,K.
It is currently the oldest observed relic of the early Universe, and its slight inhomogeneities \citep{Smoot1992} allow us to reconstruct the initial density perturbations which, via
gravitational collapse, evolved into the cosmic structures observable today \citep{White1978}.

After recombination, the Universe entered an era of darkness devoid of visible light, known as the cosmic dark ages. During this period, matter overdensities were gradually amplified due to their enhanced gravitational attraction and eventually catalyzed the formation of the first stars around redshift 30 \citep{Klessen2023}.
These so-called Population~III (Pop~III) stars were metal-free and emitted copious amounts of Lyman continuum photons (LyC; $\geq 13.6\,\mathrm{eV}$) capable of ionizing hydrogen in their surroundings.
The first stars also polluted their ambient, pristine interstellar medium (ISM) with metals at the end of their lifetimes, which altered the composition of subsequent generations of stars.
As these overdensities continued to evolve, they led to the formation of the first galaxies \citep{Bromm2011}.
The combined radiation from these galaxies is believed to have played a significant role in the patchy reionization of the intergalactic medium \citep[IGM;][]{shapiro1986cosmological,haardt1995radiative,gnedin1997reionization,madau1999radiative,gnedin2000effect}, creating expanding ionized bubbles that started to merge and eventually coalesced to fill the entire Universe. The exact epoch when the Universe became fully ionized is still under debate, although current observations \citep{becker2015evidence,bosman2022hydrogen} favour a later timeline than the previous paradigm of $z=6$ from \cite{fan2006constraining}.

The Epoch of Reionization (EoR) offers a unique chance to better understand the evolution from the relatively uniform density distribution observed in the CMB to the complex large-scale structure we see today.
However, observing galaxies during the EoR is challenging due to the enormous distances of these objects and their stellar spectra being redshifted to infrared wavelengths. The recently launched \textit{James Webb Space Telescope} \citep[\textit{JWST};][]{rigby2023science} has opened new avenues for observing the first galaxies, offering insights that are beginning to challenge our existing models of the EoR. Early observations are potentially finding significantly more massive galaxies at high redshifts ($z\approx 13$) than previously expected, sparking strong debate within the scientific community \citep{curtis2023spectroscopic,robertson2023identification}.
Furthermore, the growth and distribution of ionized bubbles during the EoR will be aided by future observations of 21\,cm line emission from neutral hydrogen. This method promises to become increasingly feasible with upcoming instruments such as
the Hydrogen Epoch of Reionization Array (HERA), NenuFAR, and the Square Kilometre Array (SKA).

To fully exploit the upcoming observations, it is essential that theoretical models achieve sufficient physical fidelity to make accurate predictions and reliable interpretations.
The complexity of non-linear galaxy formation physics \citep{Vogelsberger2020} often necessitates numerical simulations, although analytical methods can also yield significant insights \citep{dekel2023efficient}.
Specifically, to study the EoR, it is essential to incorporate a non-homogeneous radiation field \citep{borrow2023thesan,Shen2024}, which in turn requires sophisticated radiative transfer (RT).
While it is possible to apply RT calculations in post-processing \citep{ciardi2003simulating, iliev2007kinetic, mcquinn2007morphology, mcquinn2009he}, radiation-hydrodynamics (RHD) simulations that realistically model individual galaxies and therefore self-consistently account for the feedback of radiation onto the baryonic matter offer the most precise framework for comparison.
However, the rapid propagation of radiation combined with the need to resolve small-scale star-forming regions within galaxies as sources of reionization makes such simulations computationally expensive. 
As a result, simulations are typically constrained to relatively modest volumes 
\citep{CROC2014,codai,codaii,codaiii,SPHINX2018,SPHINX2022,Obelisk2021,Aurora2017, Finlator2018,xu2016galaxy,wells2022, bhagwat2023spice}. This limitation underscores the ongoing challenge in the field --- balancing a detailed treatment of (sub-)galactic-scale processes with the expansive spatial scales pertinent to reionization.

The {\small THESAN} simulation project \citep{Thesan1, ThesanAaron, ThesanEnrico} combined the IllustrisTNG galaxy formation model, an update to the previous Illustris galaxy formation model \citep{IllustrisModel2013,IllustrisNature,IllustrisIntro,genelIllustris}, that was successfully tested at lower redshifts \citep{Springel2018, Marinacci2018, Naiman2018, Nelson2018, Pillepich2018Model, Pillepich2018, TNGPublicDataRelease} with a novel radiative transfer module \citep{arepoRT} in the moving mesh code {\small AREPO} \citep{springel2010pur, weinberger2020arepo}.
Despite employing strategies such as the reduced speed of light approximation and subcycling (where multiple RT steps are performed per hydrodynamic integration step), the flagship simulation containing $2100^3$ hydrodynamic resolution elements was judiciously limited to only run down to redshift $z=5.5$ with a box size of $95.5\,\mathrm{cMpc}$.
To accurately capture cosmic reionization on larger scales, especially the coalescence of the largest ionizing bubbles, significantly larger boxes \citep[$>250\mathrm{cMpc}$;][]{kaur2020minimum} are required. Although {\small AREPO} simulations within the MillenniumTNG project have recently surpassed these scales, managing $4320^3$ cells run to redshift $z=0$ \citep{MTNG_main, Kannan2023},
they did not include RT physics \citep[for even larger hydrodynamics simulations, see ][]{astrid1,astridNi,flamingo}.
Running such ambitious large-volume simulations with RT would require significantly more computing power, presenting a substantial challenge for the current version of {\small AREPO-RT}.

With the launch of the Frontier supercomputer in 2022 \citep{atchley2023frontier}, we have officially entered the era of exascale computing.
By definition, these machines are capable of executing over $10^{18}$ floating-point operations per second (FLOPS) with double precision, a significant leap from previous generations of supercomputers.
In the near future, even more powerful machines will come online, including El Capitan, JUPITER, and industry-backed supercomputers \citep{chang2023simulations}.
Efficient use of such large systems offers the potential for unprecedented simulation accuracy, which presents unique challenges for general-purpose simulation codes.
These systems utilize hundreds of thousands of CPU cores alongside specialized accelerators, such as graphics processing units (GPUs), organized into computing nodes connected over a network.
Leveraging specialized libraries, such as those implementing the widely-supported Message Passing Interface (MPI), allows tasks running on different computing cores and nodes to communicate by exchanging data packages.
However, the scale of these systems introduces significant communication overhead, primarily due to the sheer volume of concurrent messages.
This bottleneck can be mitigated by exploiting shared main memory within computing nodes, allowing for direct memory access for intra-node communication and consolidating messages between nodes.
These optimizations can be implemented, for example, by employing the MPI-3 shared memory API or the multithreading library OpenMP.
Given Frontier's current configuration of 64 cores per node \citep{atchley2023frontier}, and with future systems expected to exceed 100 cores per node, such strategies are becoming even more critical.

Another challenge is that most of the computational power in these exascale machines comes from the GPU-acceleration.
Specialized programming languages or frameworks are required to harness their full parallel processing capabilities. This adjustment often involves algorithmic modifications and memory layout optimization to minimize latency and data transfer between CPUs and GPUs.

Despite these advancements, only a few astrophysical codes are currently optimized for GPU utilization, and they are often specifically designed for GPU clusters.
Examples of the latter type are {\small CHOLLA} \citep{cholla2015,cholla2024}, {\small TENET-GPU} \citep{Cernetic2023, cernetic2024supersonic}, {\small ATON} \citep{Aubert2010}, {\small RAMSES-CUDATON} \citep{codai,codaii}, and {\small  K-ATHENA} \citep{KAthena}. However, they are usually restricted to run only on uniform spatial grids, which significantly simplifies the efficient use of GPUs.
A recent notable exception is {\small AthenaPK}, which has ported some modules of the {\small ATHENA++} code \citep{athenapp} to GPUs on an AMR grid using the {\small PANTHEON} framework \citep{pantheon2022}.
Similarly, the {\small P-GADGET 3} code \citep{Springel2005} has implemented GPU-accelerated gravity and smoothed-particle hydrodynamics (SPH) for an unstructured resolution element distribution.
It takes advantage of the OpenACC compiler extensions and achieves speed-ups between 2 and 4 \citep{Ragagnin2020}.

The primary goal of this paper is to optimize and enhance the moment-based RT solver from \cite{arepoRT} and enable it to make full use of the exascale machines mentioned above. This entails exploiting the shared memory within nodes to reduce communication costs and offloading the expensive RT calculations to GPU accelerators.
The RT module is particularly well-suited for such an optimization since the unstructured mesh is static during subcycles, which minimizes the amount of data that must be copied to the GPU per calculation. RT also typically dominates the total computational load in cosmological RT simulations, such as those performed by the {\small THESAN} project.
Therefore, these optimizations have the potential to substantially reduce simulation run times. Alternatively, they could allow more accurate treatments of the radiation, for example, by increasing the number of frequency bins or transport physics.

This paper is structured as follows:
In Section~\ref{sec:RTOnMovingMesh}, we introduce the moving-mesh methodology employed by {\small AREPO} with a specific focus on the original CPU-centric implementation of the momentum-based RT solver from \cite{arepoRT}. We concentrate in this section on the algorithmic structure of the module, which is used throughout the rest of the paper. Utilizing the {\small THESAN} simulation as a case study, we demonstrate the necessity of accelerating both the communications between MPI tasks and the computations themselves to maximize the efficiency of RT in cosmological simulations.
We introduce in Section~\ref{sec:newCommunication} an innovative communication strategy that uses shared memory to replace MPI communications within each node with direct memory accesses and to bundle multiple inter-node messages into a single message, significantly reducing the number of MPI calls.
Through simulations of an expanding \HII region, we demonstrate the superior scalability of this new scheme compared to the old one.
In Section~\ref{sec:RTOnGPU}, we discuss a new implementation of RT for GPUs, utilizing the CUDA framework to accelerate computations.
By categorizing calculations into those requiring external data and those that do not, we effectively hide the communication overhead between CPUs and GPUs. 
We show that to use GPUs optimally, there must be enough independent work items, requiring around 100,000 Voronoi cells per MPI task.
In Section~\ref{sec:cosmologicalBox}, we apply our new methods to the more realistic case of a cosmological box simulation at redshift $z \approx 5.5$ using the {\small THESAN} model in a RHD context.
We find a considerable performance improvement for the GPU RT module compared to the original CPU module.
Finally, in Section~\ref{sec:dicussion}, we discuss current bottlenecks in our simulations and identify opportunities for further optimizations in the future.
We conclude the paper with a summary of our findings and contributions in Section~\ref{sec:conclusions}.

\section{Radiative transfer on a moving mesh}
\label{sec:RTOnMovingMesh}
The moving mesh code {\small AREPO} \citep{springel2010pur, weinberger2020arepo} solves the Euler equations that describe the evolution of a perfect fluid on an unstructured Voronoi mesh using the finite volume method.
The Voronoi mesh is constructed from a set of mesh-generating points that can move with arbitrary velocities. However, they are typically set to the local fluid velocity, resulting in a quasi-Lagrangian behaviour.
The code utilizes the Message Passing Interface (MPI) to efficiently parallelize the computation, resulting in good scaling up to about 100,000 computing cores \citep{MTNG_main}.
This is achieved by dividing the computational box into smaller domains, which are then distributed among MPI tasks.
Each task can construct its portion of the Voronoi mesh by importing cells from neighbouring domains as ghost particles, for a valid representation across domain boundaries.
{\small AREPO} has been extended in the past to solve additional equations, including self-gravity, ideal magnetohydrodynamics \citep[MHD, ][]{pakmor2011magnetohydrodynamics, pakmor2013simulations}, non-ideal MHD \citep{nonidealMHD1,nonidealMHD2}, thermal conduction \citep{kannan2016accurately}, viscosity \citep{munoz2013}, and radiative transfer \citep[RT;][]{arepoRT, Jaura2018, jaura2020sprai, arepoMCRT, Peter2023}.
The basis for our momentum-based RT solver is the original CPU implementation described in \cite{arepoRT}. {\small AREPO-RT} employs a second-order accurate operator splitting scheme, which allows us to discuss the RT implementation independently of the hydrodynamics.

\subsection{Moving mesh method in {\small AREPO}}
The Euler equations are fundamental in describing the conservation laws of mass, momentum, and energy in fluid dynamics. The equations are presented as follows:
\begin{equation}
    \frac{\partial \bm U}{\partial t} + \nabla \bm{\cdot} \bm F(\bm U) = 0 \, .
    \label{eq:eulerDifferential}
\end{equation}
 The vector of primitive quantities $\bm U$ and the flux function $\bm F$, are given by,
\begin{equation}
\bm U  = \begin{pmatrix}
   \rho \\
   \rho \bm v  \\
   \rho e  \\
   \end{pmatrix} \quad \text{and} \quad
   F(\bm U) =  \begin{pmatrix}
   \rho \bm v\\
   \rho \bm v \bm v^T + P\\
   \rho e  \bm v + P \bm v\\
   \end{pmatrix} \, ,
   \label{eq:eulerFlux}
\end{equation}
where $\rho$, $\bm v$, $e$, $P$ are the density, velocity, total energy per mass, and pressure, respectively. The total energy density $e = u + \frac{1}{2} \bm v^2$ consists of the thermal energy  per mass $u$ and the kinetic energy density $\frac{1}{2} \bm v^2$. 
The system of equations is closed by the ideal gas equation of state, $P = (\gamma - 1) \rho u$, with $\gamma$ as the adiabatic index relating the pressure to other thermodynamical quantities.

Equation~(\ref{eq:eulerDifferential}) can be discretized on an arbitrary space-filling mesh by integrating it over the volume $V_i$ of cell $i$:
\begin{equation}
    \frac{{\rm d} \bm Q_i}{{\rm d} t} = -\int_{\partial V_i} \bm F(\bm U) \cdot \text{d} \bm{A} \, ,
\end{equation}
where $\bm Q$ is the vector of conserved quantities
   $\bm Q_i = \int_{V_i} \bm U \text{d}V$
and we applied Gauss law. 
The flux function describes the exchange of these quantities through the cell surface $\partial V_i$, with $\bm{{A}}$ the area vector.
The average flux between two cells is calculated with the HLLD Riemann solver \citep{miyoshi2005multi} that takes into account the vectors of primitive variables of cell $i$ and its neighbours.
To achieve second-order accuracy in spatial resolution, we calculate the gradients of $\bm U$ using a least square fit \citep{pakmor2016improving} and linearly extrapolate the primitive variables to the interface.
The time integral is approximated by a second-order accurate scheme that is a hybrid between the Runge–Kutta method and the MUSCL–Hancock scheme \citep{pakmor2016improving}.
For a more detailed discussion of the used approximations and the generalization to the moving mesh, we also refer to \cite{zier2022simulating}.

To achieve numerical stability, the time step of each cell is constrained by the von Neumann stability condition:
\begin{equation}
   \Delta t_\ts{hydro} \leq \eta \frac{\Delta x}{c_s} \, ,
 \label{eq:timestephydro}
\end{equation}
where $\Delta x$ is the effective cell width, $c_s$ the sound speed, and $\eta \approx 0.3$ the Courant factor.
In large simulations, $\Delta t_\ts{hydro}$ can vary over several orders of magnitude, and {\small AREPO} therefore integrates cells with an individual time step rather than a global time step.
The individual time steps are mapped to a power-of-two time step hierarchy for efficiency. 
Each task constructs the Voronoi mesh only for its active cells and calculates the fluxes for interfaces with at least one active neighbour.
For passive particles, only their conserved quantities are updated, taking into account the fluxes from the interfaces they share with active cells.

\subsection{Radiative transfer with the M1 closure}
The radiation field can be fully described by the specific intensity ${I_\nu({\bm x},
t, {\bm n}, \nu)}$, at position ${\bm x}$ and time $t$,  as the rate of
radiation energy $E_\nu$ flowing per unit area d${\bm A}$, in the direction
${\bm n}$, per unit time d$t$, per unit frequency interval d$\nu$
centered on frequency $\nu$ and per unit solid angle d$\Omega$,
  $\text{d}E_\nu = I_\nu({\bm x}, t, {\bm n}, \nu)  \ ({\bm n} \bm{\cdot} \text{d}{\bm A})\, \text{d}t \, \text{d}\nu \, \text{d}\Omega \, $.
The evolution of $I_\nu$ can be described by the continuity equation \citep{mihalas1999foundations}
\begin{equation}
  \frac{1}{c} \frac{\partial {I_\nu}}{\partial t} + {\bm n} \bm{\cdot}  {\bm \nabla} I_\nu =  j_\nu - \kappa_\nu \,  \rho \, I_\nu \, ,
\label{eq:RT}
\end{equation}
where we introduced the emission term $j_\nu$ and the absorption coefficient $\kappa_\nu$.
Solving the full continuity equation becomes prohibitively expensive for large numbers of sources due to the high dimensionality of the problem that not only requires discretization in time and space but also in angular
and frequency variables.
Instead, \cite{arepoRT} uses a moment-based method that treats radiation as a photon fluid following the equations:
\begin{equation}
  \frac{\partial E_r}{\partial t} + {\bm \nabla} \bm{\cdot} {\bm F_r} = S - \kappa_\ts{E} \, \rho \, {\tilde c} \, E_r \, ,
\label{eq:E}
\end{equation}
\begin{equation}
  \frac{\partial {\bm F_r}}{\partial t} + {\tilde c}^2 {\bm \nabla} \bm{\cdot} {\mathbb P_r} =  - \kappa_\ts{F} \, \rho \, {\tilde c} \, {\bm F_r}\, ,
\label{eq:F}
\end{equation}
where the radiation energy density $E_r$, flux ${\bm F_r}$, and pressure
${\mathbb P_r}$ are defined as
\begin{equation}
\{{\tilde c}E_r,  \ {\bm F_r},  \  {\tilde c}{\mathbb P_r}\} = \int_{\nu_1}^{\nu_2}\int_{4\pi} \{1, {\bm n}, {\bm n}\otimes{\bm n}\}I_\nu  \, \text{d}\Omega \, \text{d}\nu \, .
\end{equation}
$S$ denotes the source term which
quantifies the amount of radiation energy emitted, $\kappa_\ts{E}$ and $\kappa_\ts{F}$ are
the radiation energy density and radiation flux weighted mean opacities within
the frequency range defined by ${[\nu_1, \nu_2]}$ and $\rho$ is the density of
gas in the cell. 
${\tilde c}$ is the signal speed of radiation transport, which under the reduced speed of light approximation (RSLA) can be smaller than the actual value.
To solve equations (\ref{eq:E}) and (\ref{eq:F}) we employ the Eddington closure relation
\begin{equation}
{\mathbb P_r} = E_r \, {\mathbb D} \, .
\label{eq:edd}
\end{equation} 
where ${\mathbb D}$ is the Eddington tensor encoding the direction of photon flux at any point.
As described in \cite{arepoRT}, we adopt the M1 closure to express ${\mathbb D}$ as a function of $ E_r $ and $\bm F_r$, which provides a practical local expression that is computationally efficient even for a large number of sources.
See \cite{thomas2022comparing} for a detailed comparison of M1 with other approximate closure relations.

The photon transport terms on the left side of equations (\ref{eq:E}) and (\ref{eq:F}) can be cast into the form of (\ref{eq:eulerDifferential}) with the vector of primitive variables $\bm U_r$, of conserved quantities $\bm Q_r$ and flux $\bm F$:
\begin{equation}
\bm U_r  = \begin{pmatrix}
   E_r \\
   \bm F_r  \\
   \end{pmatrix}, \;\;\;\;
   \bm Q_r  = \int_V\begin{pmatrix}
   E_r \\
   \bm F_r  \\
   \end{pmatrix}, \;\;\;\;
   \bm F(\bm U) =  \begin{pmatrix}
   \bm F_r \\
   {\tilde c}^2 {\mathbb P_r}\\
   \end{pmatrix} \, .
   \label{eq:RT_flux}
\end{equation}
Consequently, the same numerical methods used for solving the Euler equations can also be used to simulate the evolution of the photon fluid, with the exception that no rest frame exists for the photon fluid.

Finally, the source terms in the RT equations can be solved independently of the photon transport using Strang operator splitting.
They can also exchange energy with interacting gas and dust through photon absorption and emission, which depends on the fluid's chemical structure.
In the rest of the paper, we use the nonequilibrium chemical network described in \cite{arepoRT} that includes hydrogen and helium ionization.
As a consequence, we generally discretize the photon energy spectrum into three bins to model the ionization of \HI, \HeI, and \HeII.

\subsection{Algorithmic structure of radiative transfer in {\small AREPO}}
\begin{figure*}
    \centering
    \includegraphics[width=1\linewidth]{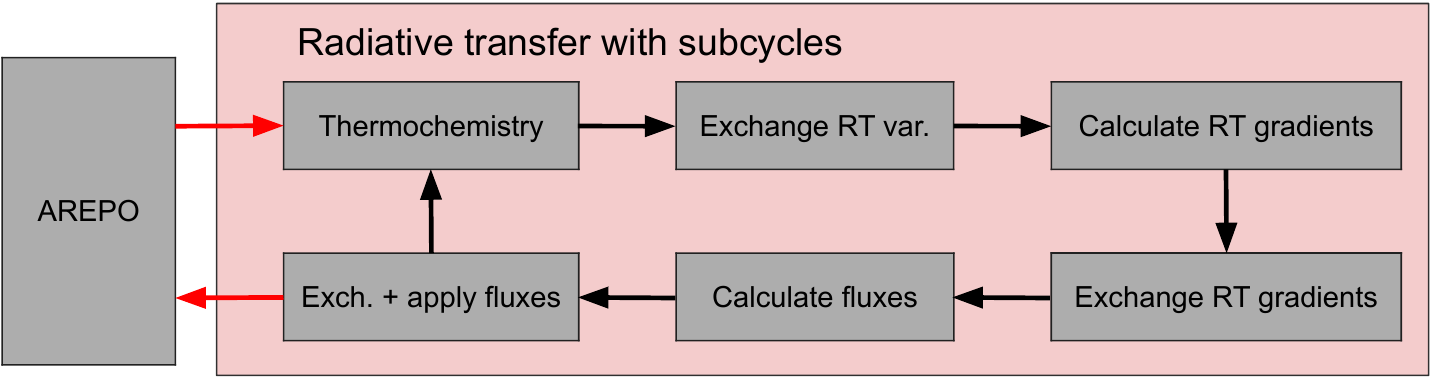}
    \caption{Overview of the substeps performed in the RT solver. We first calculate the new thermochemical state, exchange the new primitive variables with neighbours, calculate the gradients of the RT variables, exchange the gradients with our neighbours, calculate fluxes between neighbours, and exchange the fluxes and apply them. If we use subcycling, the circle formed by the black arrows will be traversed several times before leaving the module with the red arrow.}
    \label{fig:overview_subcycling}
\end{figure*}
As discussed in \cite{arepoRT}, the time step of the photon transport equation has to fulfill the von Neumann stability condition:
\begin{equation}
 \Delta t_\ts{RT} \leq \eta \frac{\Delta x}{\tilde{c} + \left|{\bm v}_{c}\right|} \, ,
 \label{eq:timestep}
\end{equation}
where ${\bm v}_{c}$ represents the velocity of the Voronoi cell in the lab frame, $\tilde{c}$ the (reduced) speed of light, and $\eta \propto 0.3$ the Courant number.
The final time step of each cell is given by:
\begin{equation}
 \Delta t = \text{min}\left(\Delta t_\ts{RT}, \Delta t_\ts{hydro}, \Delta t_\ts{grav}\right), 
\end{equation}
where we use the hydrodnamic time step $\Delta t_\ts{hydro}$ defined in equation~(\ref{eq:timestephydro}) and the gravitational time step $\Delta t_\ts{grav}$. Since the speed of light is typically significantly larger than the speed of sound, adding the RT equations drastically increases computational costs. To overcome this, the RSLA is often adopted when characteristic velocities are significantly smaller than the speed of light. However, this is not necessarily the case in cosmological reionization simulations, where the expansion rate of the ionization fronts (I-fronts) in the IGM can be close to the speed of light \citep{Rosdahl2013, Bauer2015}.
One option to decrease the simulation time is to use subcycling of the RT equations. This involves performing $N_{\rm sub}$ RT steps per hydrodynamic step, allowing for more frequent updates of the radiative state without repeatedly recalculating gravitational forces or reconstructing the Voronoi mesh.
However, this method is only effective if the RT calculations do not dominate the computational costs, as the
expensive operations of calculating gravitational forces and constructing the Voronoi mesh are performed less frequently.
Since the mesh remains static during subcycles, we are be able to optimize the code for GPUs more easily, as we will discuss later.
Typically, a value of e.g. $N_{\rm sub} = 64$ is used to balance accuracy and efficiency \citep{Thesan1}.
In \cref{fig:overview_subcycling}, we give an overview of the structure of the RT module with subcycling, and we will discuss the different submodules in detail in the following.

\subsubsection{Thermochemistry}
The thermochemistry submodule calculates the coupling between photons and gas.
The module consists of a single loop over all active cells, with independent calculations for different Voronoi cells.
The inputs include the radiation field, temperature, density, and initial chemical abundances, and the module updates these state parameters to reflect changes due to photon interactions.
An explicit integration scheme for the source terms can lead to numerical instability, especially if a cell becomes ionized during the time step.
To counteract this, even smaller timestep intervals are sometimes necessary. This can be achieved by implementing an additional subcycling, as demonstrated in \cite{Rosdahl2013}. 
In contrast, \cite{arepoRT} uses a semi-implicit scheme by default, which offers a balance between stability and computational expense. This scheme implicitly updates all quantities except the thermal energy, while the latter is still integrated explicitly. 
If the thermal energy changes by more than 10\%, a more expensive, fully implicit scheme will be used.
The fully implicit scheme uses the {\small SUNDIALS} {\small CVODE} library \citep{hindmarsh2005sundials}  and is typically only necessary for a small number of cells, resulting in good computational efficiency.
We refer to \cite{arepoRT} for a complete description of the implicit and semi-implicit schemes.

\subsubsection{Gradient estimates}
To achieve high spatial accuracy in the RT flux calculations, we move from piecewise constant (first-order) to piecewise linear (second-order), which requires gradient estimates of the primitive RT variables.
We start with a scalar field $\phi$ within each Voronoi cell with the centre of mass $\bm s_i$.
Following the method described by \cite{pakmor2016improving}, the gradient $\left< \nabla \phi\right>_i$ for cell $i$ is determined by the condition
\begin{equation}
    \phi_j = \phi_i + \left< \nabla \phi\right>_i \cdot \left(\bm s_j - \bm s_i\right),
\end{equation}
which can be only fulfilled by a linear field for all neighbours $j$.
We therefore chose $\left< \nabla \phi\right>_i$ such that the deviation from this condition is minimized:
\begin{equation}
   S_{\text {tot}} = \sum_j \frac{A_{ij}}{|{\bm s_j} - {\bm s_i}|^2} \left(\phi_j - \phi_i - \left <{\bm \nabla}\phi \right >_i ({\bm s_j} - {\bm s_i})\right)^2 \, ,
\end{equation}
where we introduced the area $A_{ij}$ of the interface between cell $i$ and $j$. The first term corresponds to a weighting scheme, which emphasizes contributions from larger and closer neighbouring cells.
The gradient can be obtained by solving the normal equation:
\begin{align}
 \sum_j \frac{A_{ij}}{|{\bm s_j} - {\bm s_i}|} \left(\bm{n}_{ji} \otimes \bm{n}_{ji} \left| \bm{s}_j - \bm{s}_i \right|   \left< \nabla
  \phi \right>_i - \left( \phi_j - \phi_i \right) \bm{n}_{ji}\right)= 0
\label{eq:normalEquation}
\end{align}
with $\bm{n}_{ji} = \left( \bm{s}_j - \bm{s}_i \right) /
\left| \bm{s}_j - \bm{s}_i \right|$, which can be done by inverting the matrix in the first term. This matrix only depends on the mesh geometry and, therefore, can be reused between subcycles of the same time step.
To calculate the gradient, we therefore only have to calculate in each subcycle the second term of equation~(\ref{eq:normalEquation}), which can be implemented as a loop over all neighbours and requires the updated variables $\phi_j$ from the thermochemistry.

To suppress numerical instabilities, we have to use a slope-limiter that prevents the creation of new extrema by linear extrapolation. To achieve this, we calculate the minimum and maximum of $\phi_j$ and compare those values with those we obtain in the middle of each interface of cell $i$ if we use linear extrapolation. If the value is larger (smaller) than $\text{max}(\phi_j)$ ($\text{min}(\phi_j)$), we reduce $\left< \nabla \phi\right>_i$ by a constant factor to prevent the formation of the new extremum.
The gradient calculation is implemented as a loop over all active cells. 
For each cell, we iterate over all its neighbours to calculate the second term of equation~(\ref{eq:normalEquation}) and save $\text{max}(\phi_j)$ and $\text{min}(\phi_j)$. We multiply by the pre-computed inverse of the matrix from equation~(\ref{eq:normalEquation}), and iterate again over all neighbours to apply the slope-limiting, ensuring the gradients do not introduce non-physical behaviour.

\subsubsection{Flux calculation}
\label{subsubsec:oldFluxCalculation}
The calculation of the RT fluxes in equation~(\ref{eq:RT_flux}) is performed individually for each interface that is adjacent to an active cell. 
At first, the gradients are used to perform a linear extrapolation of the primitive variables to the centre of each interface. Optionally, a time extrapolation using the linearized RT equations can also be applied to predict future states. However, this is incompatible with subcycling and individual time steps and is therefore not employed in this paper. 
To simplify the geometry, the new primitive variables at the interface are rotated into a coordinate system where the normal vector of the interface aligns with the $x$-axis, to be passed into the Riemann solver. 
The resulting fluxes are rotated back into the lab frame, and a further limiter is applied to prevent negative photon densities.
The final flux is applied to the local cells, modifying the conserved quantities based on the net flux across each interface.
We note that for interfaces formed by two cells lying on different MPI tasks, the flux over the interface is calculated only once, and the result is communicated to the other task.
Upon receiving the flux data, the imported values are applied to the local cells, ensuring photon conservation.
 
\subsection{Computational costs in the {\small THESAN} simulations}
\label{subsec:compCostsThesan}
\begin{figure}
    \centering
    \includegraphics[width=1\linewidth]{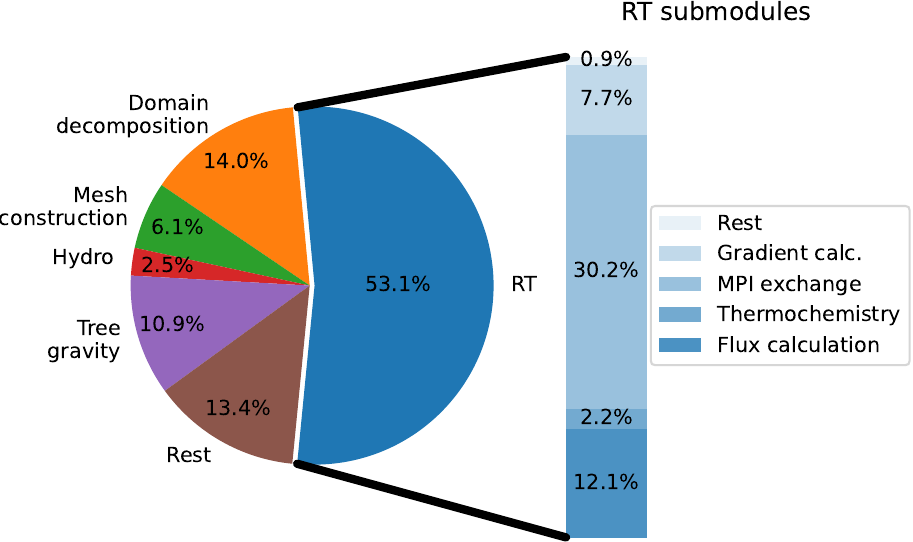}
    \caption{The distribution of the total run time of the {\small THESAN}-1 simulation over the different submodules. The RT solver dominates the computational costs, which is dominated by the communication costs between MPI tasks.}
    \label{fig:thesanRunTime}
\end{figure}

To better understand the computational costs of the different submodules in the RT solver, we analyze the run time distribution from the flagship simulation {\small THESAN}-1 of the {\small THESAN} project \citep{Thesan1, ThesanAaron, ThesanEnrico}.
It evolves a representative volume of the Universe ($95.5 \, \text{cMpc}$ on a side) from redshift $z = 49$ to $z = 5.5$ using $2100^3$ dark matter particles and approximately an equal number of Voronoi cells.
The simulation required $\approx 28$\,million CPU-hours across 57,600 cores on the SuperMUC-NG machine at the Leibniz-Rechenzentrum, which provides compute nodes with two Intel Xeon Platinum 8174 processors, summing up to 48 cores per node.
As shown in \cref{fig:thesanRunTime}, the RT solver dominates the computational cost, which implies that increasing the number of subcycles would only slightly reduce the simulation times. 
The cost of the RT solver is approximately equally split between communication costs between different MPI tasks and actual computations, suggesting that both need to be optimized to reduce the run time significantly.
Since communication overhead typically increases with the number of MPI tasks, using more computing nodes cannot effectively reduce the total run time.

\section{Exploiting shared node memory for inter-task communication}
\label{sec:newCommunication}
\begin{figure*}
    \centering
    \includegraphics[width=1\linewidth]{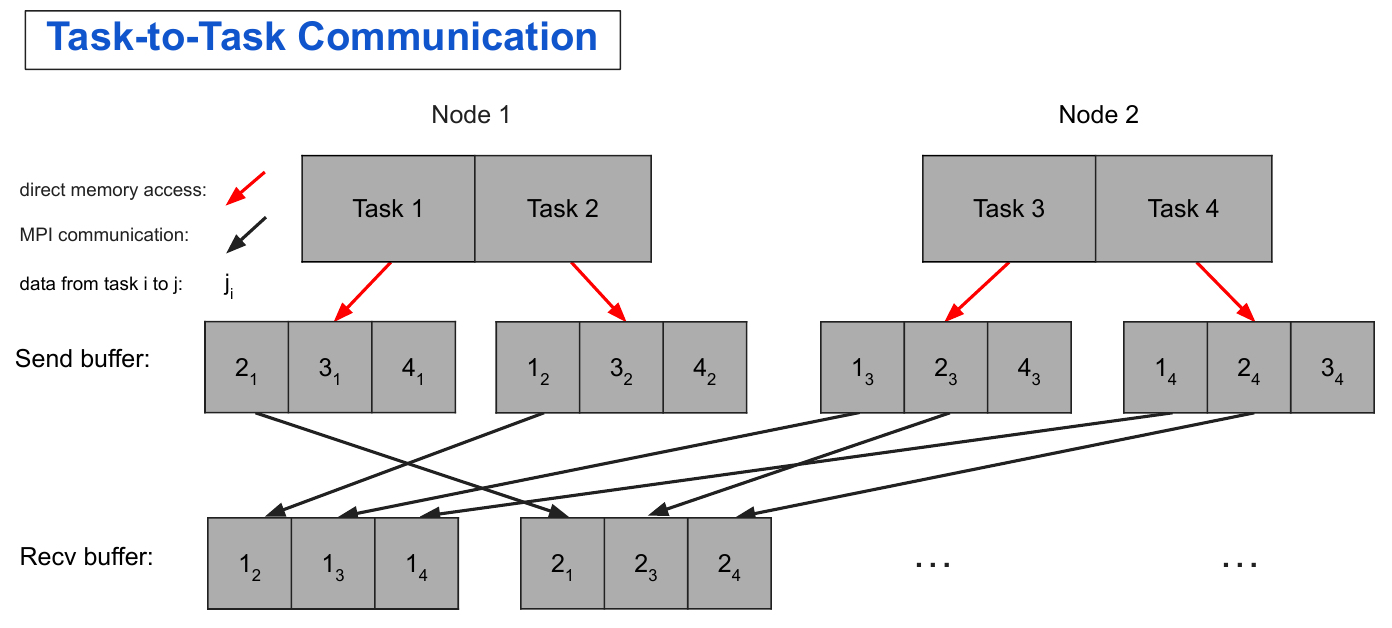}\\
    \vspace{1cm}
     \includegraphics[width=1\linewidth]{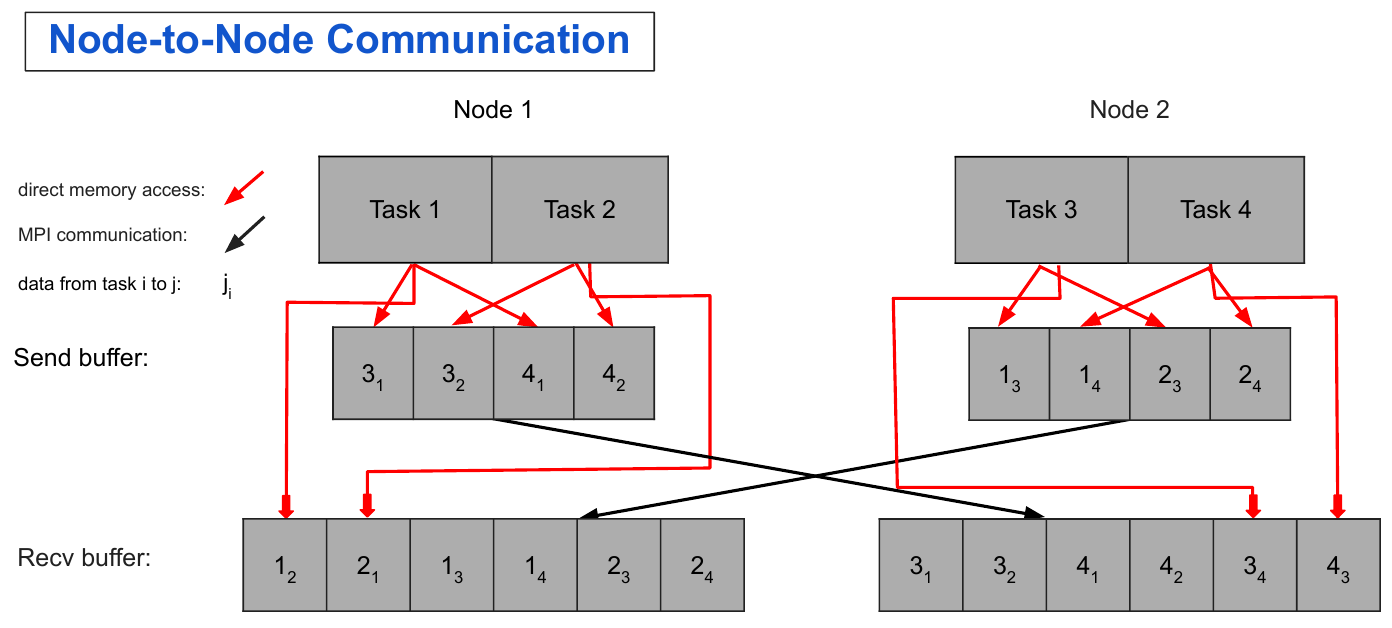}
    \caption{A comparison of the old (task-to-task) and new (node-to-node) communication schemes for the simplified case of two computing nodes with two cores each. 
    In the old scheme, each task had its own export buffer, which contained the data to be exported sorted by the MPI task number of the destination.
    Using pair-wise MPI communication, the data is exchanged and ends up in the task-specific import buffer.
    In the new scheme one export and one import buffer per node are allocated in the shared memory. 
    Data exchanged within the node can be directly written into the import buffer. In contrast, all messages to the other node can be bundled in one, reducing the number of MPI calls and allowing better use of the entire network bandwidth through larger MPI message sizes.}
    \label{fig:old_communication_scheme}
\end{figure*}
As we can see in \cref{fig:overview_subcycling}, during each subcycle in the RT solver, three sets of quantities have to be communicated: The primitive variables required for the gradient calculations, the gradients needed for the flux calculations, and finally the fluxes through the interfaces on the surface of the MPI tasks' domain required to update the conserved variables.
All this communication is ``local'', i.e., a task only needs to communicate with another task if they have neighbouring Voronoi cells in their domains.
In the original implementation, each MPI task allocates memory for an individual export buffer, which they fill with the data to be exported, sorted by destination.
They also allocate an import buffer to hold the data from the other tasks and start iterating over all other MPI tasks in the simulation.
During each iteration, the MPI tasks form communication pairs using a hypercube and blocking MPI communication to exchange data.
The call to the MPI library is omitted if the pair does not need to communicate because their domains do not contain neighbouring cells. However, much of the time must be spent waiting for the other task to finish its previous communication.
The use of non-blocking MPI communication could reduce the required synchronization, but for several thousand tasks, the performance is degraded by a large number of pending MPI calls, which makes it infeasible for large cosmological simulations. In \cref{fig:old_communication_scheme}, we sketch the communication scheme for the simplified case of two nodes with two MPI tasks.

This paper introduces a new communication scheme called node-to-node communication, which differs from the previous task-to-task communication strategy.
This scheme is based on the trend that the number of computing cores per computing node has significantly increased over the last decade, with new machines having over 60 cores per node \citep[e.g., $N=64$ for Frontier,][]{atchley2023frontier}.
The cores within each node share the same main memory. Using the MPI 3.0 shared memory API, tasks operating on the same node can directly write to and read from this shared memory.
Therefore, explicit communication within one node with MPI messages is unnecessary, and MPI tasks within one node can share one export and import buffer.
The latter allows bundling messages from all $N$ tasks on node $i$ to all $N$ tasks on node $j$ into a single MPI message. This optimizes communication bandwidth utilization and reduces the impact of network latencies.
Since it significantly reduces the total number of individual MPI messages by $\mathcal{O}(N^2)$, it also helps to push the MPI library's scalability limit to much larger simulations. This could allow asynchronous communication calls for medium to large simulations.

In practice, we begin by synchronizing all tasks within a node and allocating shared export and import buffers, which can be reused between subcycles. 
Each task writes data for other nodes into the export buffer, sorted by the destination's MPI task number. It is important to note that this requires task numbering such that consecutive task numbers are on the same node, which can be enforced.
The tasks write data to the shared import buffer when they export it to another task on the same node.
To ensure the export buffer is ready, we must synchronize the tasks within the node before iterating over all other nodes. 
During each iteration, we create pairs of computing nodes to communicate using a hypercube with blocking communication calls.
Each MPI task within the node communicates with every $N^\text{th}$ node, allowing simultaneous communication with $N$ nodes at a time.
After communication, the tasks within each node must be synchronized to ensure the validity of the import buffer.
In \cref{fig:old_communication_scheme}, we present the new communication scheme for the simplified case of two nodes with two tasks.

\subsection{Scaling for simulating an expanding \HII region}
\label{subsec:expandingHIIRegionCommunication}
\begin{figure}
    \centering
    \includegraphics[width=1\linewidth]{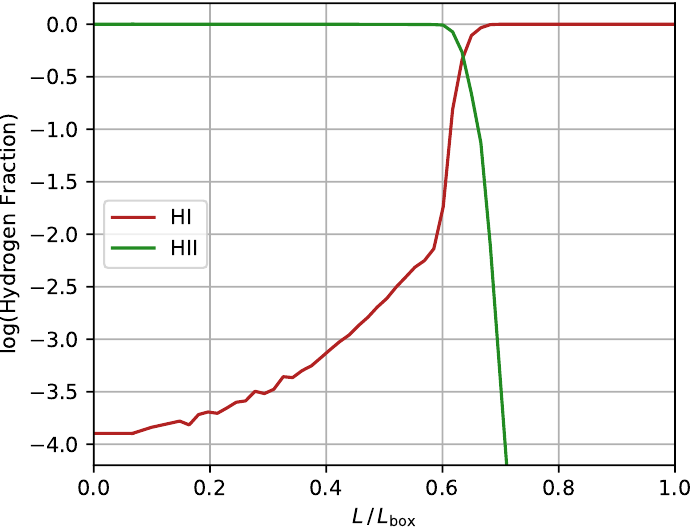}
    \caption{ The spherically averaged ionization profile of a single \HII region at $t=25\,\mathrm{Myr}$. 
    In this case, we use only the hydrogen chemistry to compare the run time of our new communication scheme with the old one. 
    For more details, see Section~\ref{subsec:expandingHIIRegionCommunication}.}
    \label{fig:hii_region_profile}
\end{figure}
\begin{figure}
    \centering
    \includegraphics[width=1\linewidth]{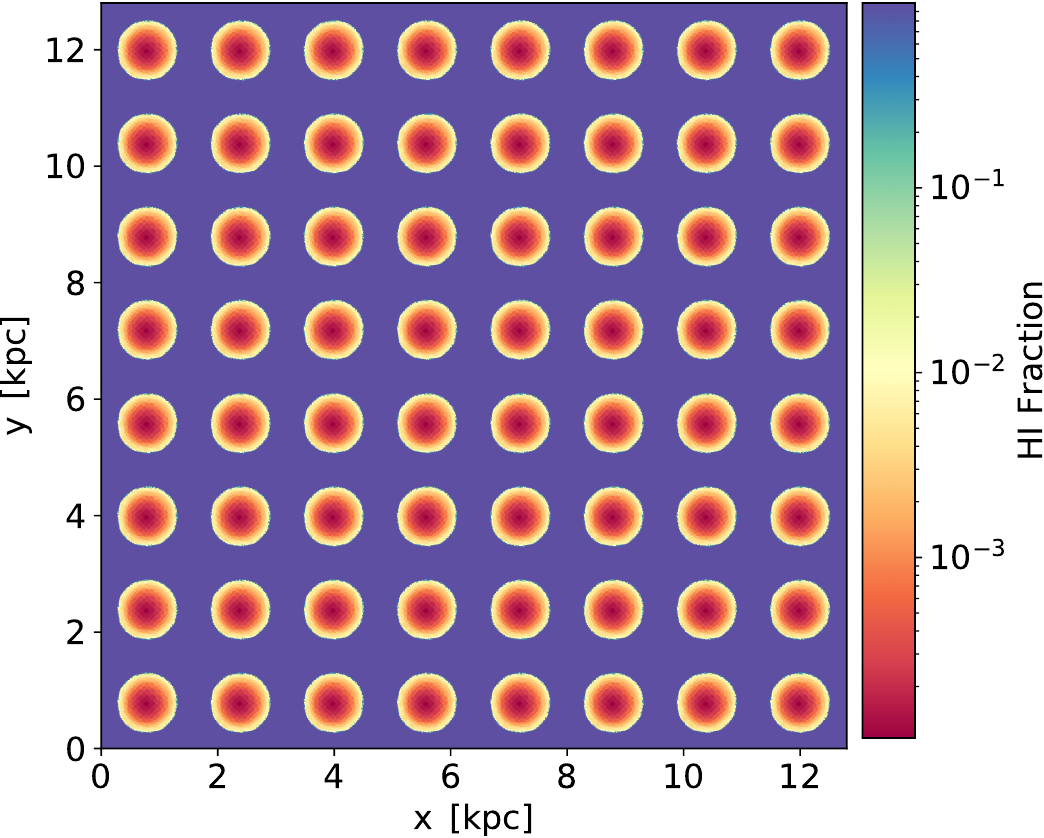}
    \caption{A slice through a simulation box containing 512 \HII regions, which were simulated for the weak scaling test with 256 nodes described in Section~\ref{subsec:expandingHIIRegionCommunication}. We show the distribution of the \HI fraction at $z =0.8\mathrm{kpc}$ at $t=25\,\mathrm{Myr}$. The slice intersects with 64 \HII regions.  }
    \label{fig:hii_region_many_sources}
\end{figure}
\begin{figure}
    \centering
    \includegraphics[width=1\linewidth]{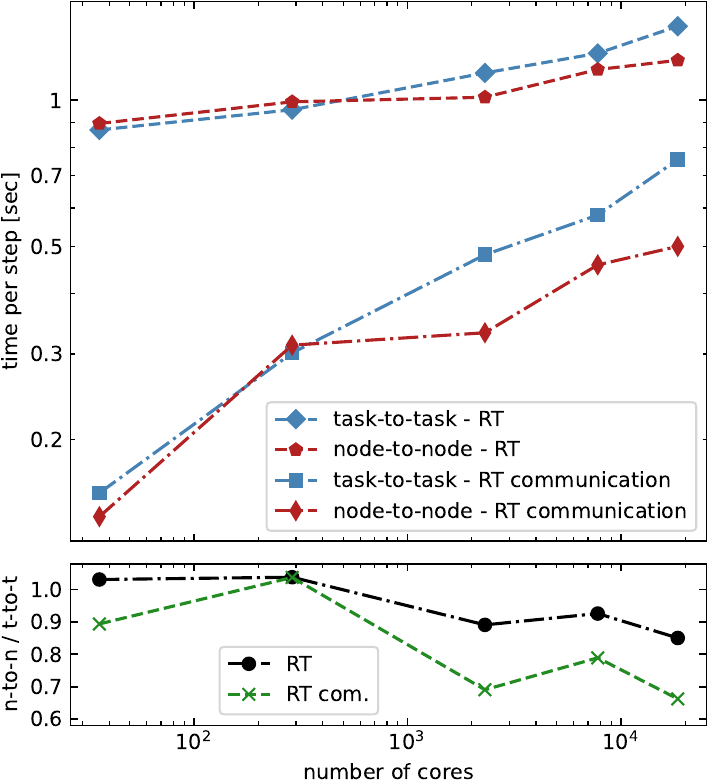}
    \caption{ \textbf{Task-to-Task vs Node-to-Node communication, weak scaling}: The average run time per time step as a function of cores for simulating an expanding \HII region on the ``Raven'' supercomputer.
    We increase the number of cells with increasing number of cores to perform a weak scaling test and run the simulation with the new (n-to-n) and the old (t-to-t) communication schemes.
    In the top panel, we show the run time spent in the RT solver and the costs of the MPI communication in the RT solver. Additionally, we show the relative run time difference for both measurements in the bottom panel.
    For more details, see Section~\ref{subsec:expandingHIIRegionCommunication}. }
    \label{fig:weakScaling_HII_raven}
\end{figure}
\begin{figure}
    \centering
    \includegraphics[width=1\linewidth]{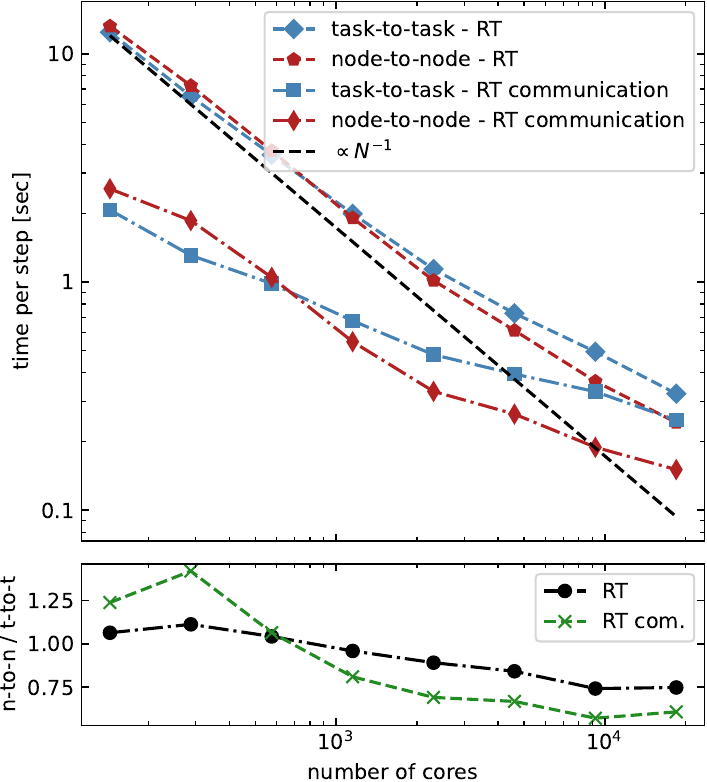}
    \caption{\textbf{Task-to-Task vs Node-to-Node communication, strong scaling}: The same as \cref{fig:weakScaling_HII_raven} but for strong scaling. We simulate 64 \HII regions and change the number of computing cores used in the simulations.}
    \label{fig:strongScaling_HII_raven}
\end{figure}
To assess the performance of our new node-to-node communication scheme, we conducted both weak and strong scaling tests using the scenario of expanding \HII regions.
The simulation setup was adapted from Section~4.5 of \cite{arepoRT}, and for simplicity, focusing on a simplified model involving a single hydrogen ionizing frequency bin.
We set up a box of size $(1.6\,\mathrm{kpc})^3$ containing pure neutral hydrogen of temperature $T = 100$\,K and a number density 
\begin{equation}
    n_H \left(r\right) = \begin{cases}
n_0 &\mathrm{if\ r < r_0}\\
n_0 \left(r_0 /r\right)^2 &\text{otherwise}
\end{cases}  
\end{equation}
with core density ${n_0=3.2\,\text{cm}^{-3}}$ and a radius of
${r_0=91.5\,\text{pc}}$.
The central source is a black body spectrum with
${T_\ts{eff}=10^5\,\text{K}}$ and emits at a rate of ${10^{50}\,\text{photons
s}^{-1}}$.
We set up a regular staggered grid with $2 \times 44^3$ cells and simulate ${25\,\text{Myr}}$ with a reduced speed of light of $10^{-3} c$ and 64 subcycles.
As shown in \cref{fig:hii_region_profile}, the medium around the source becomes ionized within a spherical \HII region.

For the scaling tests, we use the ``Raven'' supercomputer operated by the Max Planck Computing and Data Facility (MPCDF), which offers two Intel Xeon IceLake Platinum 8360Y CPUs per computing node, each with 36 compute cores.
We simultaneously simulate several \HII regions' evolution for the weak and strong scaling tests. 
We keep the standard setup described above and add periodic copies to construct larger initial conditions.
For the weak scaling, we use one source per $1/2$ computing node and increase the number of sources proportionally to the number of computing nodes.
This leads to 4,732 cells per MPI task on average.
As an example for this larger box, we show a slice through a computational box with 512 sources in \cref{fig:hii_region_many_sources}.
In \cref{fig:weakScaling_HII_raven}, we show the results of the weak scaling test up to 18,432 cores.
For small simulations, both communication schemes give similar results. However, for $>1000$ cores, the new scheme scales significantly better, though we note that both versions show good parallel performance of 61\% (old version) and 74\% (new version) in the largest simulation.
We show in \cref{fig:strongScaling_HII_raven} additionally the results of a strong scaling test, where we use in all simulations 64 \HII regions.
Both versions show good scaling, with 42.5\% (node-to-node) and 29.9\% (task-to-task) parallel efficiency, even when the number of tasks is increased by a factor of 128.
For a small number of nodes, the task-to-task communication is faster than the new one, which shows better performance for a larger number of nodes.
The additional node-internal synchronization before the communication starts for the node-to-node scheme can explain the speed difference for small simulations.
The latency costs and the MPI library's overhead increase for larger simulations, making the new scheme more efficient.
We expect this to become even more dramatic if we go to a higher core number, though this is beyond the scope of this paper.

In general, the scalability depends on the quality of the network connecting the computing nodes, both in the software and hardware.
We repeated this study on the ``Engaging'' cluster at MIT as an example of a typical university cluster, which offers only 1/4 of the network bandwidth.
As shown in Appendix~\ref{app:scalingEngaging}, the old scheme offers significantly worse scalability than the new one on this machine.

\section{Optimizing computations using GPUs}
\label{sec:RTOnGPU}
\begin{figure*}
    \centering
    \includegraphics[width=0.9\linewidth]{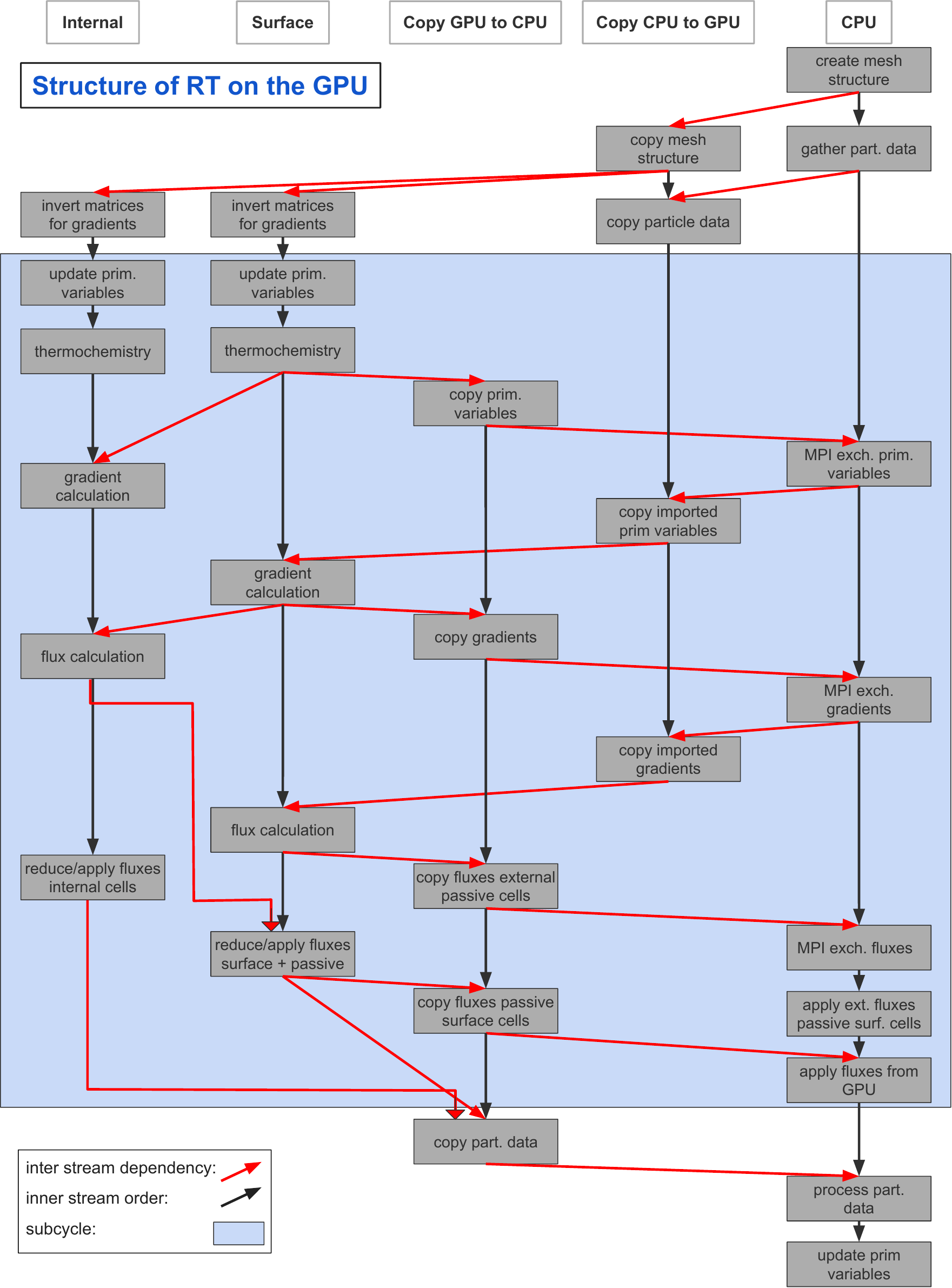}
    \caption{Dependency graph containing the main operations in the RT solver using the GPU. The four different streams on the GPU and the CPU can execute operations in parallel and are synchronized using an event system (red arrows). The operations within a subcycle are repeated several times (blue background). Operations within the same stream are automatically executed in order (black arrows). }
    \label{fig:flowDiagramGPU}
\end{figure*}
We now turn to optimizing the RT workflow under a heterogenous CPU/GPU paradigm.
As shown in \cref{fig:thesanRunTime}, in addition to MPI communication, a significant fraction of the run time is also spent on flux calculation, gradient estimation, and thermochemistry. The primary submodules of the RT solver consist of loops over the active Voronoi cells (thermochemistry and gradient calculations) or cell interfaces (flux calculations). 
These independent operations can be performed in parallel, making the RT module ideal to run on GPUs, which are optimized to execute hundreds of thousands of simultaneous threads.
In contrast to CPU programming, which allows the development of highly portable code, GPU programming requires the use of vendor-specific language extensions like CUDA (NVIDIA), HIP (AMD), or SYCL (INTEL) or frameworks that automatically generate hardware-specific executables, such as KOKKOS \citep{edwards2014kokkos} or  RAJA \citep{beckingsale2019raja}.
Our implementation only utilizes basic GPU functionalities, which allows our application to be easily ported from one vendor-specific language extension to another.
Throughout the rest of the paper, we will concentrate on our CUDA implementation but note that a similar version exists for HIP, demonstrating the adaptability to different GPU environments.

The GPU architecture cannot access the CPU main memory directly, and requires data transfers between CPU and GPU memory.
However, this relatively expensive operation can be efficiently managed by overlapping it with GPU computations and copying data from the GPU to the CPU. This concurrency on the GPU can be implemented using multiple ``streams'' per MPI task, representing independent queues of operations.
CUDA facilitates this with a relatively lightweight event system that enables synchronization between different streams.
By default, every MPI task would require its own CUDA context, which is memory-intensive and restricts the GPU to being used by only one task at a time.
Instead, the Multi-Process Service (MPS) offers a binary-compatible implementation of the CUDA API that allows tasks to share a single context and run simultaneously on the GPU, which is crucial for our application.
Functions that should be run on the GPU have to be written as a ``kernel'' that distributes the computations across multiple threads, which are organized into groups called warps consisting of 32 threads.
Threads within a warp can only execute the same operation on different data simultaneously, in a Single Instruction, Multiple Threads (SIMT) fashion. 
If a thread should not execute an operation due to a control structure such as an ``if statement'', it will sleep instead, reducing the total maximum computing throughput. This behaviour is called ``warp divergence'' and should be avoided.

Memory management is also crucial, as the threads within one warp should consecutively access memory.
Threads store local data in registers, and while using more registers can speed up access to data, it also reduces the maximum number of threads that can run concurrently on the GPU.
This reduced occupancy does not automatically lead to lower performance, but it can reduce the GPU's ability to hide memory access latencies by alternating between threads.
Furthermore, each thread has a limit of 255 registers.
Exceeding this limit forces threads to use slower memory (register spilling), which can lead to a significant slowdown of the application.
Splitting larger kernels into smaller ones can be helpful to prevent register spilling.
For best throughput, expensive synchronizations between threads should be minimized, and the number of independent operations that can run in parallel on the GPU should be maximized.

\begin{figure*}
    \centering
    \includegraphics[width=1\linewidth]{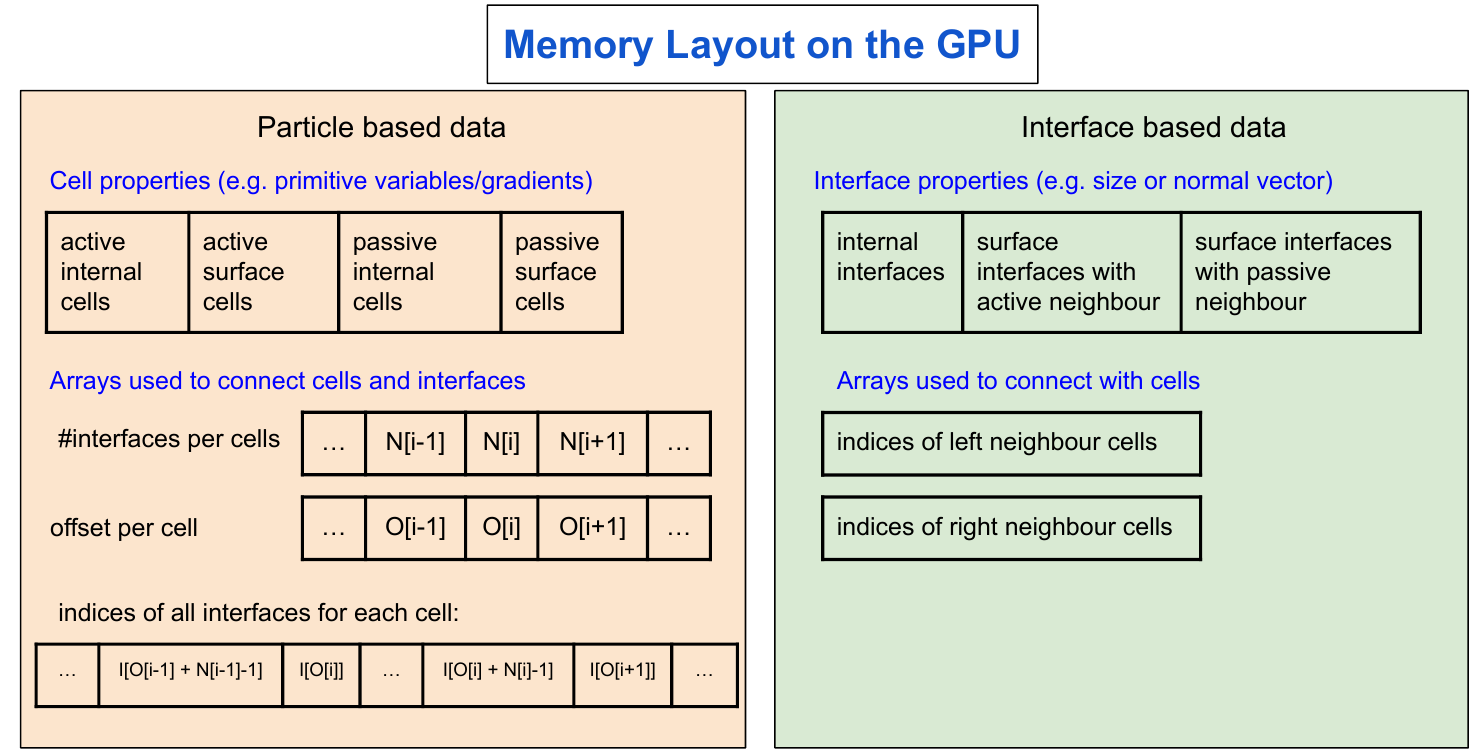}
    \caption{We show the data structures we use on the GPU for the radiative transfer.
    Properties of the particles are stored in standard arrays, which are sorted depending on whether the cells are on the surface or fully within the MPI task's domain and whether they are active or passive. 
    Similarly, properties of the active interfaces are stored in arrays, which are sorted depending on whether both neighbouring cells are on the same MPI task and whether the external neighbours are passive. In the latter case, the flux has to be exported to the CPU. 
    The sorting allows only the necessary fraction of data to be copied during the subcycles between the GPU and CPU.
    We also save a list that contains the indices of all active interfaces per cell.
    The required entries for a particular cell can be calculated using an offset and the number of interfaces per cell, which are stored in two separate arrays.
    For each interface, we also save the left and right cell indices.
    This allows us to load the required data for the flux calculations and indirectly loop for each Voronoi cell over all its neighbours, which is equivalent to looping over all its active interfaces. }
    \label{fig:GPU_memory_layout}
\end{figure*}
\subsection{Radiative transfer on the GPU}
Although one could split the computational workload between the CPU and GPU, we realized that the GPU was typically significantly faster than the CPU, so we made a strategic decision to perform all calculations on the GPU. The CPU is only responsible for the MPI communication, for which we use the new node-to-node scheme introduced in the previous section. 
We equally distribute the GPUs per node between the MPI tasks, which independently communicate with their assigned GPU.
This simplifies the GPU implementation significantly, although we note that there might be further optimization opportunities, particularly when processes using the same GPU share a boundary between their domains.

Given that the mesh geometry remains static throughout each subcycle, the set of active cells and interfaces does not change.
As a consequence, most of the data can be transferred to the GPU once before the first subcycle and only copied back after the completion of the full RT step.
Only data that needs to be exported to other tasks must be copied to the CPU in each subcycle.
To manage the workflow, we utilize four CUDA streams per task.
Two streams are responsible for copying data between the GPU and CPU. Another stream is responsible for performing calculations for cells and interfaces that do not have external neighbours.
The remaining stream performs the calculations for cells and interfaces on the surface of the task`s domain.
Splitting the calculations naturally leads to overlapping the MPI communication with calculations using only local data from the task.
To optimize this setup further, we increase the computational priority of the surface stream to ensure timely processing of boundary data.
Stream operations are coordinated using CUDA’s event system, which is necessary for tasks like the internal gradient calculation that depends on the updated primitive variables at the domain boundaries.
To illustrate the operations within this code architecture, we provide a flow chart in \cref{fig:flowDiagramGPU}, detailing each stream's activities and their interdependencies.
The following sections describe the implementation of the different submodules and the memory layout on the GPU in more detail.

\subsubsection{Thermochemistry  on the GPU}
The thermochemistry module requires updating the conserved quantities of all active cells.
The calculations for each active cell are independent; therefore, we use one thread per cell.
Although the SUNDIALS library supports GPUs, it cannot be called by individual threads, which is why we instead use the semi-implicit method from \cite{arepoRT} and additionally subcycle the thermochemistry time step if the 10\% rule for the thermal energy is violated.
To stabilize the helium chemistry, we also have to enforce the condition that the relative chemical abundances can only change by 10\% during each integration step.
In practice, we start with the initial time step $\Delta t_{\rm sub}$ of the subcycle and calculate the maximum fraction $f$ by which the thermal energy or relative chemical abundances change.
For $f > 0.1$, the step is rejected, and its size is reduced by the factor of $f_{\rm s} / f$ with typically $f_{\rm s}=0.05$.
This choice keeps the relative changes at around 5\% per integration step on average and reduces the number of rejections.
If the time integration step is accepted, we also multiply the new time step with $f_{\rm s} / f$ but enforce that the time step can, at maximum, double between two subsequent integration steps.
Although our new scheme potentially requires a significant amount of integration steps, it does not require access to the main memory within the iterations, allowing maximum computing throughput. 
One drawback of our approach is the potential for warp divergence, and sorting the particles according to the number of integration steps would be ideal. We did not find a good way to estimate this number and therefore accept warp divergence as an unavoidable part of our scheme.

\subsubsection{Gradient estimates on the GPU}
Calculating gradients for each cell is an independent operation but requires significant memory to store the maximum and minimum value of each primitive variable of each neighbour and the vector in the second term of the equation~(\ref{eq:normalEquation}).
When handling multiple photon bins (e.g. $\geq 3$) the memory requirement for each cell exceeds the 255-register limit per thread, forcing the local data to spill into slower memory. This leads to a significant slowdown of gradient calculation as the number of photon bins increases.
We decided, therefore, to use one thread per photon bin per cell and additionally to save the prefactors in the second term of the equation~(\ref{eq:normalEquation}) for each neighbour interaction, which reduces the number of redundant calculations.
Each thread performs an iteration over the neighbouring cells to calculate the sum in the second term of equation~(\ref{eq:normalEquation}), multiplies the result with the stored inverted matrix from the first term of this equation, and iterates again over all neighbours to limit the gradient.

\subsubsection{Flux calculation  on the GPU}
We use one thread per interface to calculate the fluxes between neighbouring cells.
In contrast to the CPU implementation presented in Section~\ref{subsubsec:oldFluxCalculation} we do not directly apply the results to the conserved quantities, since this could lead to a race condition, but instead save them to memory.
In an additional kernel that is parallelized over the Voronoi cells, we iterate for each cell over all of its active interfaces and apply the saved fluxes.
This allows us to avoid any additional synchronization between individual threads.
Since the flux calculations are very efficient on the GPU, fluxes for interfaces with cells from different MPI tasks are calculated independently on each task. Consequently, the resulting flux need not be copied to the CPU and communicated to other tasks.
This strategy leads to a significant speed-up during global time steps but fails if one of the cells is not active. This is because only active cells have a valid Voronoi mesh constructed on each task, and therefore, the interface is missing on the other task.
To overcome this problem, we identify these interfaces on all tasks during the initial setup and inform the task with the passive cell about it.
For these special interfaces, fluxes are still communicated in each subcycle between tasks using MPI, and updates to the conserved quantities for passive cells are managed on the CPU. This requires copying the relevant flux data from internal interfaces containing these cells from the GPU to the CPU.
Nevertheless, these interfaces are only a small fraction of the total number of interfaces shared between different MPI tasks, which reduces overall communication costs to maintain the GPU efficiency gains.

\subsubsection{Data layout on the GPU}
\label{subsubsec:dataLayoutOnGPU}
In \cref{fig:GPU_memory_layout}, we give an overview of the memory layout we use to support high-throughput operations on the GPU.
In general, we store data that does not change between subcycles in a ``structure of arrays'' (SOA) format rather than an ``array of structures'' (AOS), i.e. each property of a cell or interface is in a separate contiguous array.
This allows for more efficient memory loading on the GPU for parallel loops over all particles or interfaces.
We note that we store all primitive variables in the same structure since some have to be exchanged between the CPU and GPU each subcycle. Therefore, a full SOA implementation would require significantly more memory copies.
The same applies to the gradients.

We sort cells on the GPU into four categories, depending on whether they have an active interface with a cell on another MPI task (i.e., they are on the surface of their task's domain) and whether they are active or passive.
This simplifies splitting loops over all active cells into one over all cells with external dependencies and one over those without.
Passive cells on the surface that receive additional flux from an active interface not present on the MPI task have their conserved quantities updated on the CPU, necessitating data transfer each subcycle.
Within each category cells are sorted by their number of active interfaces to reduce warp divergence in the gradient calculation.
Similarly, we sort interfaces depending on whether both neighbouring cells are on the same MPI task, and whether the potentially external cell is active or passive.
Again, loops over all active interfaces can be split into ones over the internal and surface interfaces.
For those with a passive external neighbour, the flux has to be exported to the CPU and then to the neighbouring MPI task.

We construct additional arrays to manage the indexing of interfaces and Voronoi cells.
For each interface, we save the left and right Voronoi cell indices.
For each cell, we save the number of active interfaces and corresponding offsets.
These values can be used to find all indices of the cells' active interfaces in an additional array.
These indexing arrays provide bijective mappings between cells and interfaces, which allows efficient memory access of states and loops over all neighbours of an active cell as, e.g., required for gradient calculations.
The count array is redundant since it can be recalculated from the offset array. 
However, it is still helpful in some cases, e.g., when sorting the cells according to the number of their neighbours, and it only adds a slight memory overhead.

\subsection{Comparison of CPU and GPU performance}
\begin{figure}
    \centering
    \includegraphics[width=1\linewidth]{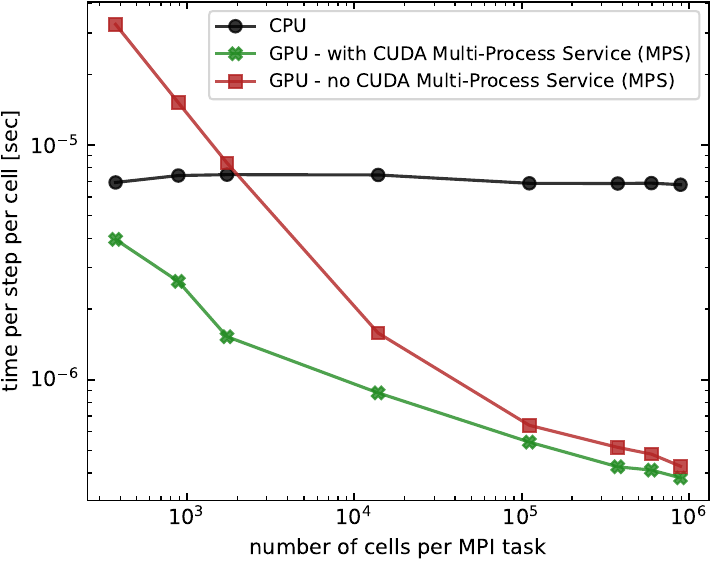}
    \caption{\textbf{CPU vs GPU, photon transport:} The run time per step per cell in seconds of the RT module in a simulation of a traveling radiative wave as a function of the number of cells per MPI task.
    We employ a static mesh and perform the simulations on one node on ``Raven'', which has 72 cores and 4 GPUs.
    We perform simulations with the old CPU and new GPU versions of RT. For the GPU implementation, we run additional simulations without the CUDA Multi-Process Service (MPS).
    The GPU requires enough independent tasks to achieve maximum performance.
    If one MPI task cannot provide enough work, MPS can allow several tasks to use the GPU concurrently.
    For more information, see Section~\ref{subsubsec:rtTravellingWave}.  } 
    \label{fig:rt_wave_gpu_vs_cpu}
\end{figure}
In this section, we validate our GPU scheme by comparing its performance against the original CPU implementation.
Since we use the same MPI communication for both versions, we expect similar scalability and will, therefore, focus on tests using a single node.
The tests were conducted on the GPU partition of the ``Raven'' supercomputer at MPCDF. Each node offers two Intel Xeon IceLake Platinum 8360Y processors for a total of 72 cores and four NVIDIA A100 GPUs.
The configuration also supports the use of the CUDA Multi-Process Service (MPS), which allows multiple MPI tasks to simultaneously utilize the same GPU.

\subsubsection{Photon transport: Travelling wave}
\label{subsubsec:rtTravellingWave}
First, we test the radiation transport in isolation, i.e., the gradient and flux calculations without the thermochemistry solver.
We, therefore, perform a similar test to the radiation wave test from Section~4.1 from \cite{arepoRT}. Unlike the original setup, we set the optical depth to zero and use a periodic, three-dimensional box with side length $L = 1$. The initial photon density is:
\begin{equation}
    E(r) = E_{\rm bg} + \epsilon \sin \left(2\pi\left(x + y +z\right)\right) \, ,
\end{equation}
where $E_{\rm bg} = 1$ is a uniform photon background density and $\epsilon = 10^{-6}$. The radiation flux points in the direction $(1,1,1)$ and has a value  $\left| \bm F_r \right| = c E_r$, with $c=1$.
We use 64 subcycles and an irregular mesh as initial conditions, meaning a Cartesian grid with a 10\%  deviation between the mesh-generating point and the mesh centroid.
We finish the simulation at $t=4$ (four full oscillations) and vary the number of cells $N$ per dimension to better understand the performance as a function of the utilization of the GPU.
For all simulations, we use the node-to-node scheme discussed in Section~\ref{sec:newCommunication}, meaning no MPI communication calls must be performed in these simulations.

In \cref{fig:rt_wave_gpu_vs_cpu}, we show the total run time of the RT solver as a function of $N$ for CPU and GPU. 
The CPU version requires around $7 \times 10^{-6}s$ per cell update, almost independent of the number of cells.
This is expected if the communication overhead is negligible since the processing in the CPU is fully serial.
In contrast, the GPU version becomes significantly more efficient with increasing workload. Its performance only starts to saturate at $\gtrsim 10^5$ cells per task, and exhibits a maximum relative speed up of $16$ compared to the CPU.
Interestingly, without MPS, the code achieves a comparable speed for large $N$, while for small $N$, this version even becomes slower than the CPU.
This demonstrates that for good GPU performance, enough independent work has to be provided.
MPS can help if one MPI task alone cannot utilize the GPU fully.
In this test, we only used one radiation bin, and we expect the lines to move to the left for simulations with more bins.

\begin{figure}
    \centering
    \includegraphics[width=1.\linewidth]{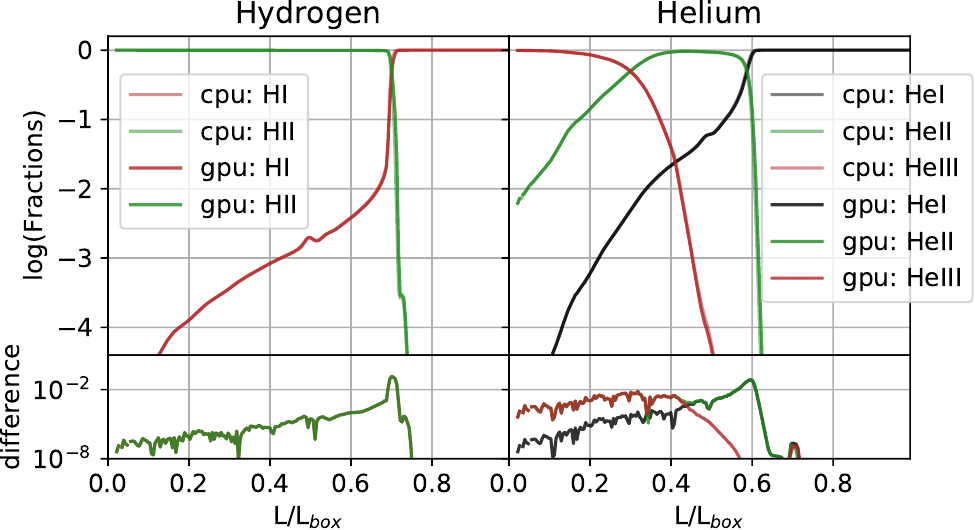}
    \caption{A comparison of the radial ionization profile of hydrogen and helium at $t=25\,\mathrm{Myr}$ in a simulation of an expanding \HII region with the CPU and GPU versions. 
    We used a resolution of $2\times 250^3$ for these simulations and show additionally the absolute differences of the ionic abundances from each simulation. Except for a tiny shift close to the ionization fronts, the results are in excellent agreement, which can be explained by the different algorithms used to integrate the chemical network.
    More details can be found in Section~\ref{subsubesc:expandingHIIRegionGPU}.}
    \label{fig:structure_gpu_HII_008}
\end{figure}

\begin{figure}
    \centering
    \includegraphics[width=1\linewidth]{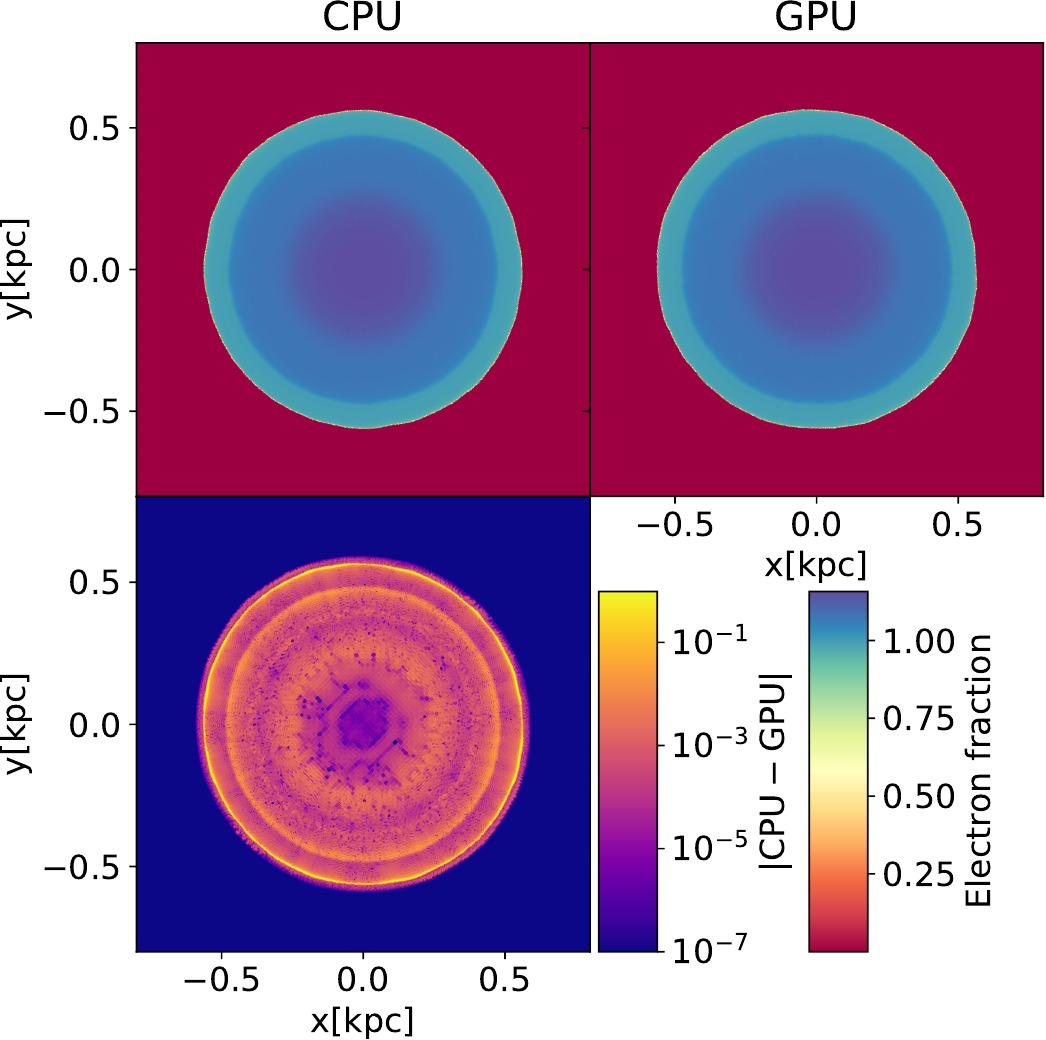}
    \caption{In the top panels, we show a slice through the centre of an \HII region containing hydrogen and helium at $t=25\,\mathrm{Myr}$ showing the free electron abundance.
    Close to the star, the electron fraction is highest since both hydrogen and helium are fully ionized.
    Further away, \HeII dominates, and finally, helium becomes neutral.
    At the outermost layer of the \HII region, only hydrogen is ionized.
    The bottom panel shows the absolute difference between the simulations run with the two code versions.
    Both agree very well for this setup with a resolution of $2\times 250^3$ cells. More details can be found in Section~\ref{subsubesc:expandingHIIRegionGPU}.}
    \label{fig:08_ne_heii_cpu_vs_gpu}
\end{figure}
\subsubsection{Coupled RT: Expanding \HII region}
\label{subsubesc:expandingHIIRegionGPU}
As a more realistic setup, we analyze a modified setup from the expanding \HII region test presented in Section~\ref{subsec:expandingHIIRegionCommunication}. 
We use the full hydrogen--helium chemical network, including three photon bins to track the ionization of \HeI and \HeII alongside \HI.
We inject the photons homogeneously within a sphere of $0.05\,\mathrm{kpc}$ radius at the centre of the box, sourcing $10^{50}$, $5\times 10^{48}$, and $3\times 10^{48}$ photons per second for each respective bin.
We assume a black body spectrum at $T=$ 50,000\,K to calculate the flux-weighted cross-sections and photoheating rates for each bin, which are essential for thermochemistry.
We use a full computing node on ``Raven'' and vary the number of cells as $2\times N^3$ to evaluate the performance under increasing workloads.

\begin{figure}
    \centering
    \includegraphics[width=1\linewidth]{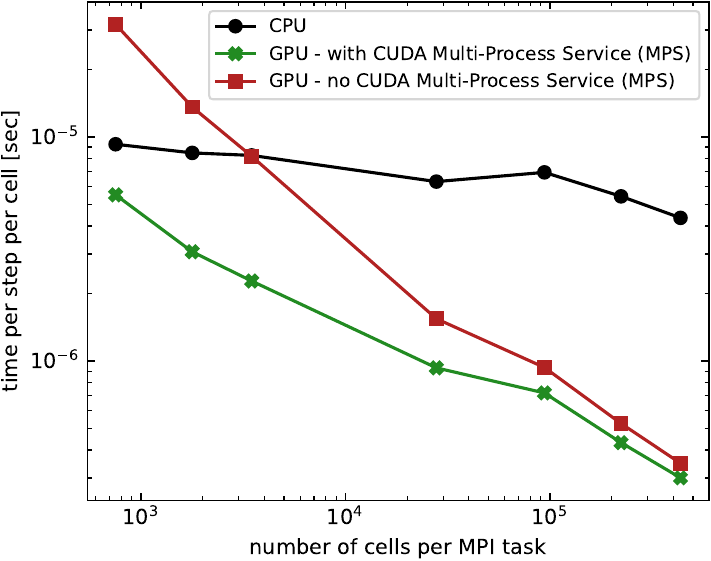}
    \caption{\textbf{CPU vs GPU, full RT in expanding \HII region:} The run time per step per cell in seconds for the RT module in a simulation of an expanding \HII region with hydrogen and helium chemistry.
    All simulations are run on one full node on ``Raven'' with 72 cores and 4 GPUs.
    We varied the resolution to measure the performance as a function of the number of cells per MPI task. We performed simulations with the old CPU and new GPU versions, with and without the CUDA Multi-Process Service (MPS). MPS significantly improves performance for low occupancy by allowing several tasks to use the GPUs concurrently.
    For more details, see Section~\ref{subsubesc:expandingHIIRegionGPU}.  }
    \label{fig:hii_gpu_vs_cpu}
\end{figure}

As we show in  \cref{fig:structure_gpu_HII_008}, the ionization profile of helium and hydrogen for the CPU and GPU implementation agree well. There are only slight deviations in the exact structure of the ionization front, which can be attributed to differences in the algorithms used to integrate the chemical network during rapid temperature changes.
As one can also see in \cref{fig:08_ne_heii_cpu_vs_gpu}, we identify three different subregions within the \HII region: near the star, where the gas is fully ionized; a middle zone where \HeII predominates; and an outer region where only hydrogen is ionized.
We also varied the resolution and measured the average run time per time step in the first $25\,\mathrm{Myr}$ for both CPU and GPU versions, with and without the use of MPS.
We note that individual time steps were employed in this setup, which means that the number of time steps slightly differs between the CPU and GPU implementations. Also, the depth of the time step hierarchy may change with $N$, affecting both accuracy and performance metrics.
We present the scaling test results in \cref{fig:hii_gpu_vs_cpu}.
As expected, the CPU implementation performance is almost independent of the problem size. 
We attribute the slight variations we did not observe in the last section to the individual time steps.
In contrast, GPU performance improves significantly with increasing problem size since the parallel architecture can be more saturated.
Utilizing MPS further increases performance, particularly when a single MPI task does not have enough work to fully utilize the GPU, continuing to observe speed-ups of around 15\% even for the largest tests.
For large problem sizes, the GPU implementation is found to be about 15 times faster than the CPU version, which agrees with the results from the last section and confirms that the inclusion of thermochemistry and individual time steps does not influence our conclusions.

\section{Reionization simulations on GPUs}
\label{sec:cosmologicalBox}
\begin{table}
    \centering
    \caption{Overview of all cosmological simulations performed for this paper. 
    We choose a box with side length $L_{\rm box}$ and evolve it between redshift $z_{\rm start}$ and  $z_{\rm end}$ using $N_{\rm nodes}$ nodes on the ``Raven'' cluster, each offering 72 cores and 4 A100 GPUs. We keep the mass resolution constant, i.e., increase the particle number $N_{\rm particles}$ with increasing volume. The initial conditions for larger boxes are created by duplicating the results of the smallest box at redshift $z=5.8$. All simulations are run with the CPU and GPU versions of RT.}
    \begin{tabular}{c|c|c|c|c|c}
    \hline
        Test & $L_{\rm box}$ [cMpc] & $N_{\rm particles}$ & $z_{\rm start}$ & $z_{\rm end}$ & $N_{\rm nodes}$\\
        \hline
        Evolution & 23.3 & $2\times 256^3$ & 49 & 5 & 1\\
        \hline
        & 23.3 & $2\times 256^3$ & 5.8 & 5.7 & 1\\
        Weak & 46.6 & $2\times 512^3$ & 5.8 & 5.7 & 8\\
        Scaling & 69.9 & $2\times 768^3$ & 5.8 & 5.7 & 27\\
        & 93.1 & $2\times 1024^3$ & 5.8 & 5.7 & 64\\
        \hline
        & 46.6 & $2\times 512^3$ & 5.8 & 5.7 & 4\\
        Strong & 46.6 & $2\times 512^3$ & 5.8 & 5.7 & 16\\
        Scaling & 46.6 & $2\times 512^3$ &5.8 & 5.7 & 32\\
        & 46.6 & $2\times 512^3$ & 5.8 & 5.7 & 64\\
        \hline
    \end{tabular}
    \label{tab:overviewCosmologicalSimulations}
\end{table}

\begin{figure}
    \centering
    \includegraphics[width=1\linewidth]{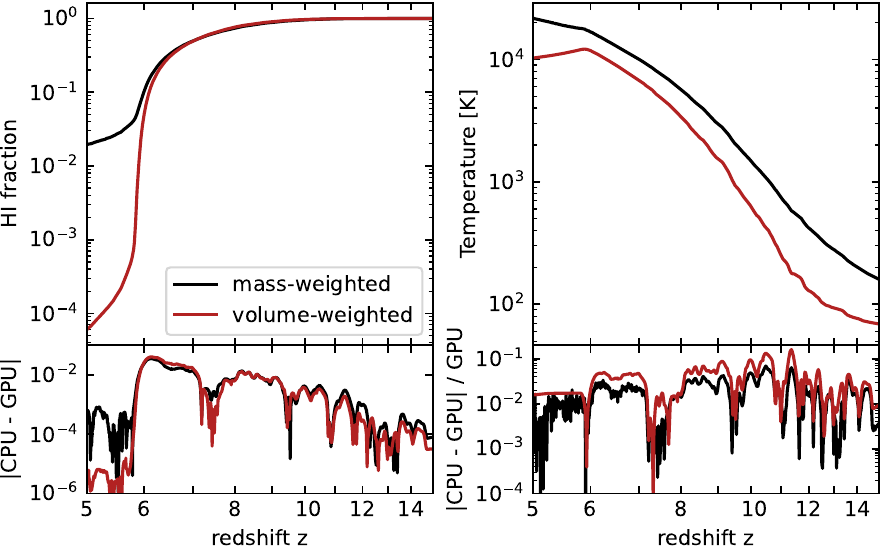}
    \caption{The evolution of the mass- and volume-weighted abundances of \HI and temperature within the standard cosmological box with side length $L_{\rm box}=22.8\mathrm{cMpc}$. We note that the \HeII abundance closely follows the \HII reionization history, with minor differences due to the presence of \HeI and \HeIII.
    At redshift $z=5$, the box is almost fully ionized except for some high-density regions, which can also be observed in \cref{fig:cosmoBoxSlice}.
    In the bottom panels we also show the differences for all quantities between the simulations run with the CPU and GPU based RT solver.
    There are small deviations, which we attribute to the different integration methods for the chemical network and intrinsic randomness in the galaxy formation model.}
    \label{fig:global_reionization_history}
\end{figure}

\begin{figure*}
    \centering
    \includegraphics[width=1\linewidth]{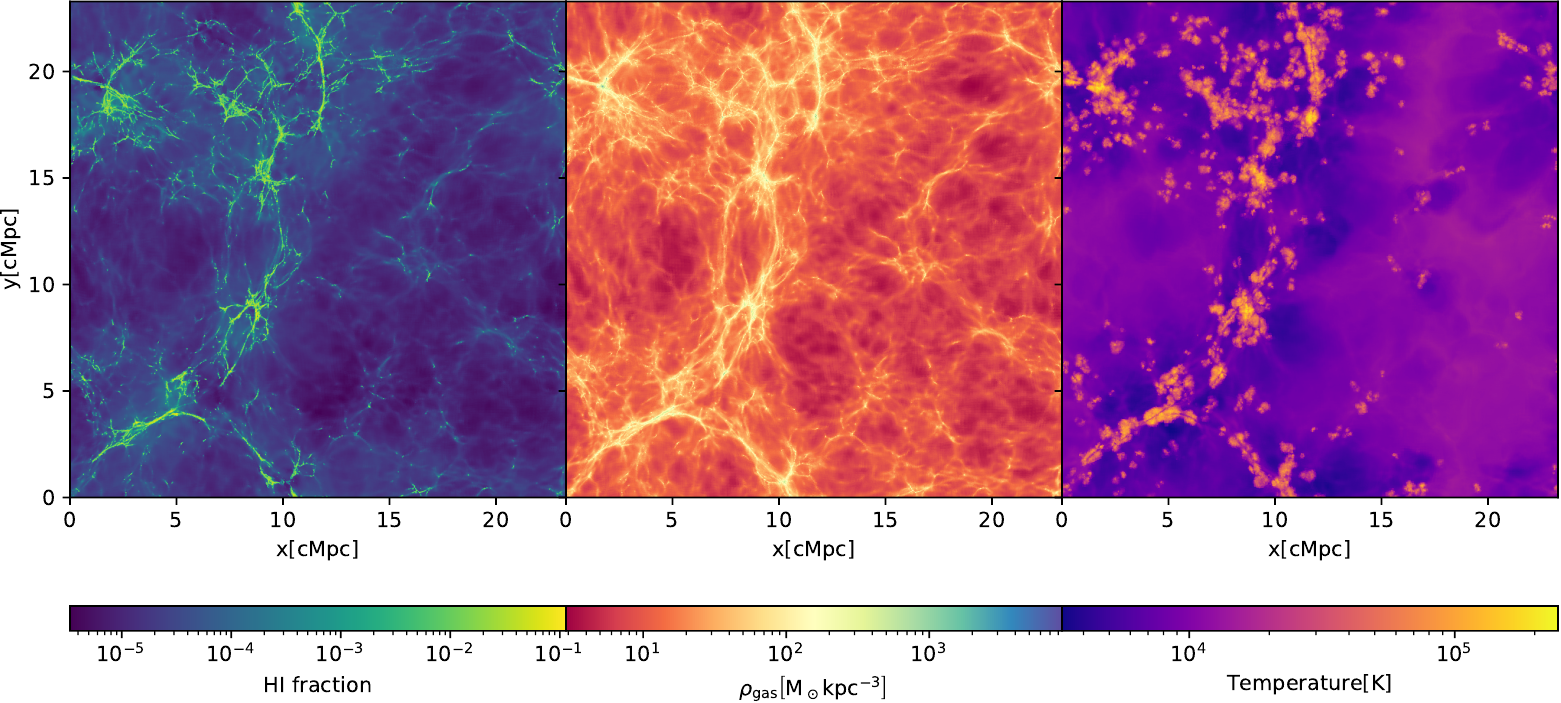}
    \caption{A projected slice through a cosmological volume at redshift $z=5.5$. 
    We average the values over a thickness of 10\% of the box size and show the neutral hydrogen fraction, temperature, and gas density. The morphology is dominated by a filament containing high-density, neutral gas across which stars and galaxies are forming.
    Feedback leads to the heating of the medium around the filament.
    The IGM contains a smooth UV radiation field and is almost isothermal, with a temperature between 10,000\,K and 20,000\,K.
    For this plot, we used the base simulation with RT on the GPU.}
    \label{fig:cosmoBoxSlice}
\end{figure*}

\begin{figure*}
    \centering
    \includegraphics[width=1\linewidth]{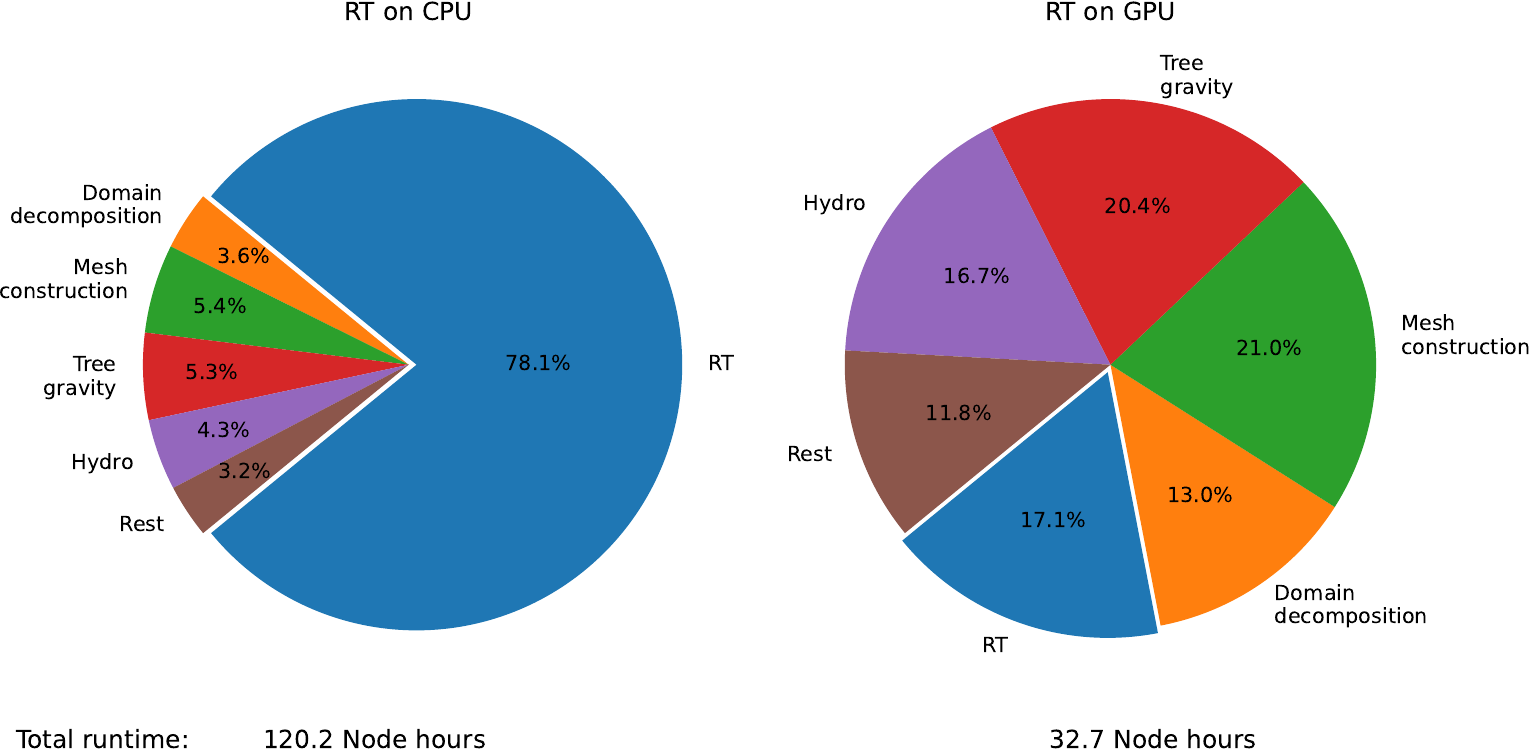}
    \caption{The distribution of the total run time for the different physics modules in the full cosmological simulations evolved from $z=49$ to $z=5$.
    In the simulation with RT on the CPU, the RT fully dominates the computational costs, while the GPU version requires a similar amount of time as the other modules.}
    \label{fig:pie_chart_gpu_vs_cpu}
\end{figure*}

\begin{figure}
    \centering
    \includegraphics[width=1\linewidth]{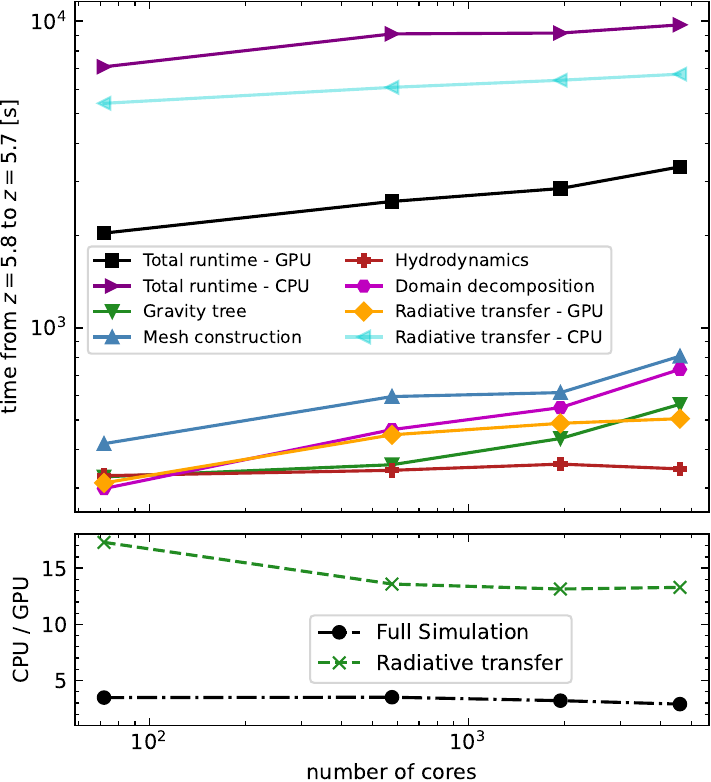}
    \caption{\textbf{Weak scaling -- Cosmological box:} Results for weak scaling tests of the CPU and GPU radiative transfer implementations for a cosmological simulation evolved from redshift $z=5.8$ to $z=5.7$.
    We keep the number of cells per node constant and show the total and RT module run times for both code versions as a function of the number of CPU cores. Additionally, we show the scaling of several other physics modules in the GPU run, which is nearly identical to the CPU run.
    In the lower panel, we show the ratio of the CPU-to-GPU run times for both the full simulation and the RT module in isolation.}
    \label{fig:weak_scaling_Cosmo}
\end{figure}

\begin{figure}
    \centering
    \includegraphics[width=1\linewidth]{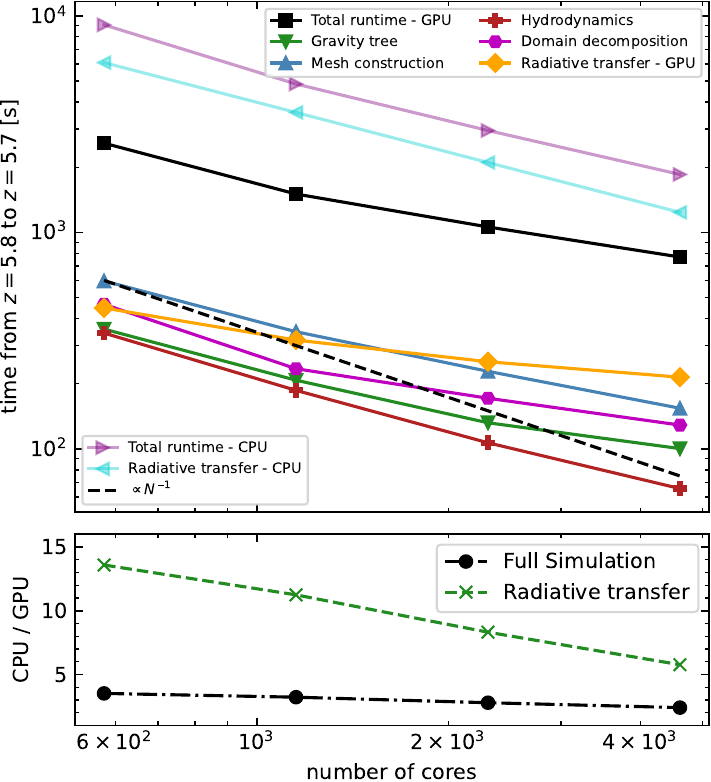}
    \caption{\textbf{Strong scaling -- Cosmological box:} Similar to \cref{fig:weak_scaling_Cosmo} but for strong scaling. In each case we run a simulation with box size $\left(45.6\,\mathrm{cMpc}\right)^3$ but vary the number of cores.}
    \label{fig:strong_scaling_Cosmo}
\end{figure}

\begin{figure}
    \centering
    \includegraphics[width=1\linewidth]{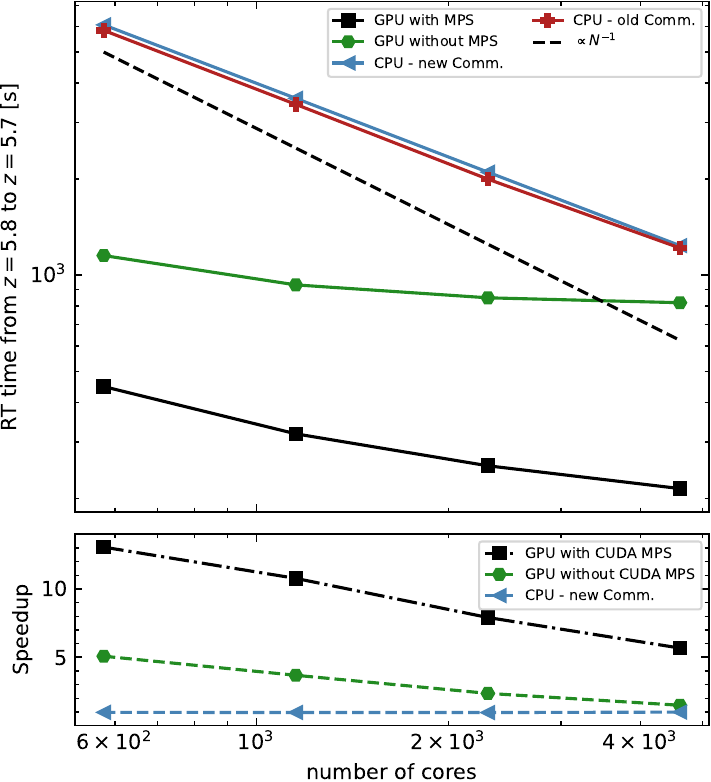}
    \caption{\textbf{Strong scaling \underline{only} RT -- Cosmological box:} Similar to  \cref{fig:strong_scaling_Cosmo}, but focusing on the strong scaling of the RT module. In addition to the preferred GPU version with the CUDA Multi-Process-Service (MPS) and the CPU version with the new node-to-node communication scheme that we already showed in \cref{fig:weak_scaling_Cosmo}, we show additional results for simulations on the GPU without MPS and simulations with the original RT module using the task-to-task communication scheme. 
    In the lower panel, we show the speedup compared to the original version of the RT module.
    MPS significantly increases performance, but the GPU scaling breaks down, since especially for small time bins, there is not enough work to use the full parallel architecture of the GPU.}
    \label{fig:strong_scaling_Cosmo_mps}
\end{figure}

To assess the performance of our GPU implementation under more complex and realistic conditions, we conduct scaling tests for simulations of the epoch of reionization, similar to the {\small THESAN} simulations.
As we have shown in Section~\ref{subsec:compCostsThesan}, for this setup, the RT dominated the computational costs, which limited the feasibility of running significantly larger boxes.
For these tests we adopted a target mass resolution of $4.66\times 10^6\,M_\odot$ for baryons and $2.49 \times 10^7\,M_\odot$ for dark matter, which is equivalent to the resolution of the {\small THESAN}-2 simulation from \cite{Thesan1}. This resolution has been shown to yield an approximately converged global reionization history for this model.
We use the same physics modules as the original {\small THESAN} simulation, including three radiation bins and the complete chemical network for hydrogen and helium. However, we also incorporate the reduced absorption scheme suggested by \cite{jaura2020sprai} and implemented in \cite{deng2024simulating}, which is beneficial for unresolved \HII regions.

We use the same cosmological parameters as \cite{Thesan1} and generate initial conditions with the {\small GADGET-4} \citep{springel2021simulating} code for a starting redshift of $z=49$.
As in the original setup, we also use 64 subcycles for the RT and a reduced speed of light of $0.2c$.
As we are primarily interested in demonstrating performance, for computational efficiency, we first evolve a relatively small box of size $22.8\,\mathrm{cMpc}$ with $256^3$ Voronoi cells down to redshift $z=5$. 
From a snapshot at $z=5.8$, we create larger boxes by periodical replication.
The modified snapshots will be further evolved to  $z=5.7$, while measuring the wall time.
By the end of the EoR, the simulated universe is already highly inhomogeneous, and the individual time steps of different cells vary by a factor of $128$.
The smallest time bin, containing only a few hundred cells, poses a particular challenge to scalability, representing a worst-case scenario for a full cosmological simulation focusing on high-redshift applications.
By replicating a smaller box, our setup does not contain larger-scale modes and is missing rare objects, which would require starting with a larger box initially. Nevertheless, we anticipate that our scaling results should remain valid, especially at comparable mass resolutions.
In \cref{tab:overviewCosmologicalSimulations}, we give an overview of all setups, which we simulated with the CPU and GPU versions of RT.
Again, we use the ``Raven'' cluster for all our simulations.

We summarize in \cref{fig:global_reionization_history} the global reionization history from the full cosmological simulation.
At the final redshift, $z=5$, the box is almost fully ionized, though some neutral gas survives in the highest-density regions within galaxies and filaments of the cosmic web.
Reionization occurs slightly later than in \cite{Thesan1}, which is typical for smaller boxes as we use here.
In contrast to a previous box size of $\left(11.4\mathrm{cMpc}\right)^3$ we experimented with, it leads to a noticeably smoother reionization history.
Some surviving neutral gas can also be seen in \cref{fig:cosmoBoxSlice}, which lies in the dense filaments that also correspond to the sites of structure and galaxy formation.
By redshift $5.5$, a smooth UV background pervades throughout the IGM, which has photoheated the gas to a few $10^4$\,K.

As we have performed the base simulation with both the CPU and GPU implementations of RT, we compare in \cref{fig:pie_chart_gpu_vs_cpu} the distribution of the total run time of both simulations onto different physics modules.
For the CPU version, around 78\% of the run time was spent on the RT, an even higher fraction than in the {\small THESAN}-1 simulation (see \cref{fig:thesanRunTime}).
As we will show later, this can be explained by the relatively good scaling of RT compared to other modules, which results in a decreasing fraction for much larger simulations.
In the simulation with RT on the GPU, only 17\% of the total run time was spent on RT, due to significantly improved computational efficiency.
The RT module was 17 times faster, which made the full simulation $3.7$ times faster.

To better understand the performance boost of moving the RT calculations to the GPU, we show in \cref{fig:weak_scaling_Cosmo} the results of the weak scaling tests we introduced before.
Since we only ported the RT algorithm, the rest of the code requires the same time in both simulations.
The computational costs in the GPU run are dominated by the mesh construction and, for larger simulations, by the domain decomposition, both of which negatively impact the scaling of the full code.
Although the RT on the GPU shows an increase in computational time when going from one node to several nodes, it otherwise scales very well.
The RT on the CPU runs 17.2 times slower than the RT on the GPU for a single node and 13 times slower for multiple nodes.
This results in a total speedup of the simulation by a factor of 3.5 for one node and around 3 for many nodes.
We note that this ratio decreases for larger simulations because the scaling for certain non-RT modules degrades for this particular setup.

In \cref{fig:strong_scaling_Cosmo}, we show additionally the results of a strong scaling test within a box of size $(\mathrm{45.6\,cMpc})^3$.
Again, the domain decomposition shows the worst scaling behaviour of the non-RT modules.
The CPU version of RT shows good, strong scaling behaviour in contrast to the GPU version.
Under this setup, the module will barely become faster, especially for $\gtrsim 10^3$ cores if more computing nodes are used.
This leads to a decreased acceleration compared to the CPU version for a large number of nodes, as we, e.g., find only a speedup of 6 for 64 nodes.
At this point, we have around 18,000 cells per MPI task, which is not enough to use the full parallel capacity of the GPU (see also \cref{fig:hii_gpu_vs_cpu}), especially for small time steps.
This becomes even more obvious in \cref{fig:strong_scaling_Cosmo_mps}, where we show additional results for simulations on the GPU but without the CUDA Multi-Process Service (MPS) and additional simulations with the old communication scheme.
Without MPS, we find a maximum speedup of 5 for the RT compared to the original CPU implementation.
The module also barely becomes faster if we increase the number of nodes since most of the GPU is idle.
The two communication schemes show similar efficiency in this setup.
For a small number of nodes, the additional synchronization of all computing cores within one node before the communication for the new scheme reduces efficiency.
For larger simulations, the new scheme scales slightly better and it is expected to continue to benefit large production runs the most.

\section{Discussion}
\label{sec:dicussion}
\subsection{Current performance bottlenecks on the GPU}
We found a significant speedup by porting our RT algorithm to GPUs, which is less than the factor of 30 \cite{grete2020k} reported for the {\small K-ATHENA} code on the SUMMIT supercomputer, but similar to the factor of 12 to 16 \cite{Cernetic2023} found for the discontinuous Galerkin code {\small TENET-GPU}.
Both of those codes leverage a regular Cartesian grid with fixed memory access patterns, which enables optimizations of the memory layout to more easily exploit the faster shared memory on the GPU.
Also, the fixed cadence of operations reduces the chance of warp divergences, and their use of global time steps ensures enough work to consistently fully saturate the GPU.
In our case, using the unstructured Voronoi mesh leads to regular uncoalesced memory accesses, particularly evident during neighbour iterations in gradient calculations and when loading particle data for the flux calculations.
Due to these random memory access patterns, the gradient calculations are bound by the memory bandwidth.
By sorting the particles according to their number of neighbours, we reduced the warp divergence significantly during the gradient calculations. Nevertheless, this issue still exists in the thermochemistry calculations if some cells require more subcycles than others.

As mentioned before, on the boundary of each MPI task's domain, the tasks have to include ghost cells from their neighbours. These cells require a copy to the CPU during each subcycle, MPI communication, or direct memory access and a copy back to the GPU at the destination.
Since several MPI tasks share one GPU, this can lead to the same GPU being the origin and destination for some data.
This becomes an unnecessary significant overhead if, such as in MillenniumTNG, a hierarchical domain decomposition is used, which means that the domains of individual MPI tasks on the same node are in close spatial proximity.

\subsection{Future improvements}
Although some of the bottlenecks we discussed before, such as random memory access, are a direct consequence of the underlying use of an unstructured grid, there is still potential to optimize our implementation further.
For example, the problem of warp divergence in the thermochemistry solver can be mitigated by exploiting the observation that cells close to equilibrium typically require only one iteration, whereas cells undergoing ionization need significantly more.
To address this, we could first perform one iteration for all cells and then flag those requiring more accurate integrations.
For the flagged cells, a new kernel with fewer threads could be launched, resulting in more resources being free to be utilized for other calculations.

Another potential optimization is to adopt single precision floating point numbers, allowing us to replace expensive software-implemented function calls with hardware-optimized intrinsics.
This would be especially useful for the thermochemistry solver, which requires frequent evaluations of the exponential or power function for the temperature dependency.
This will be even more beneficial for expanded chemical networks.
Using single precision reduces the register demand per thread, potentially increasing the occupancy, which allows for better hiding memory access latencies.
It also accelerates coalesced memory accesses and allows more data to be stored on the GPU. However, one must be careful not to reduce the accuracy of the full simulation by choosing less accurate number representations. 
Consequently, while primitive variables could be converted to single precision, conserved quantities and flux summations should remain in double precision. Special care has to be taken to avoid numerical overflows, especially in the thermochemistry solver, which requires choosing a suitable unit system.
This is especially important in cosmological simulations, in which the magnitude of the radiation field can vary over many orders of magnitude.

Constructing a single copy of the Voronoi mesh for an entire node could substantially improve scalability. This would make ghost cells from MPI tasks on the same node superfluous. 
Combined with the recently added hierarchical domain decomposition, this would significantly reduce the work on the domain's surface and increase the domain's internal work---ideal for fully leveraging GPU parallelism.
This would be especially interesting for time steps with a small number of active particles since entire galaxies could lie within the domain of one node, eliminating the need to communicate with the CPU or other nodes.
Additionally, time steps with many active time bins would also be accelerated due to a reduced amount of copying data between CPU and GPU and over MPI.

The amount of data communication could also be reduced by exploiting the fact that quantities from passive cells do not change between subcycles, which means they only have to be communicated during the first subcycle.
This becomes especially important for time steps with a small number of active cells, though in this case, the communication is typically bound by latencies and not bandwidth.
Lastly, emerging hardware technologies such as Accelerated Processing Units (APUs), like the AMD Instinct MI300, which share the same memory between CPUs and GPUs, eliminating the need for redundant memories and expensive data transfers, can increase the efficiency.

 Overall, while the GPU implementation of our RT algorithm has demonstrated significant speedups, it also highlights the complexities and limitations imposed by unstructured meshes and the associated data management challenges on modern GPU architectures. Addressing these issues will be crucial for further optimizing performance and scaling up {\small AREPO} simulations efficiently.
 It could also allow the use of more accurate RT methods.

\section{Conclusions}
\label{sec:conclusions}
In this paper we presented a new, optimized version of the moment-based radiative transfer solver in the moving mesh code {\small AREPO-RT}.
Our development efforts focused on a novel communication strategy called node-to-node communication in contrast to the previous task-to-task pattern.
This approach exploits the shared memory capabilities within computing nodes to replace MPI communication with direct memory access within the node. We also combine all inter-node messages into a singular, consolidated message.
This drastically reduces the number of MPI calls and boosts scalability, as we have demonstrated through benchmark simulations of an expanding \HII region.

Further, we ported the RT solver to GPUs using the CUDA language.
We split the calculations into those lying entirely within the MPI tasks' domain and those requiring external data. Employing CUDA streams, we efficiently overlapped both types of calculations with data transfers to and from the GPU memory.
By running cosmological boxes at the end of the epoch of reionization when the density distribution is highly inhomogeneous, we showed that our new code offers excellent weak scalability up to the 4608 CPU cores and 256 GPUs we were able to access on the ``Raven'' machine. 
Notably, the GPU-enhanced version achieved a 13 times RT performance improvement over its CPU-based counterpart.

Looking ahead, to further enhance our code's performance, minimizing unnecessary communication between CPU and GPU presents a promising avenue for further gains. This can potentially be accomplished by constructing the Voronoi mesh only once per node rather than individually for each MPI task.
With these advancements, we plan to undertake simulations of considerably larger cosmological volumes during the EoR with our new GPU-based solver compared to the original {\small THESAN} project, allowing us to study the reionization of the IGM on large scales.
This increased box size will extend the predictive power of our RHD framework for upcoming 21\,cm observations, which are currently relying entirely on semi-numerical models and post-processing RT calculations. Furthermore, the additional volume is highly relevant for direct comparison with JWST results.
In particular, the formation of more massive galaxies, the impact of quasars on IGM properties, and longer sightlines for mock cosmological light cone observations relevant for Lyman-alpha forest transmission, line intensity mapping, and other observable signatures that are sensitive to the presence of large-scale bubbles.

\section*{Acknowledgements}
We extend our gratitude to Volker Springel for valuable discussions and for allowing us to use the ``Raven'' machine for our scaling tests.
We also thank Miha Cernetic for sharing insights on CUDA and GPU programming, and to R\"udiger Pakmor for useful discussions about {\small AREPO} in general.
RK acknowledges support of the Natural Sciences and Engineering Research Council of Canada (NSERC) through a Discovery Grant and a Discovery Launch Supplement, funding reference numbers RGPIN-2024-06222 and DGECR-2024-00144. MV acknowledges support through NASA ATP 19-ATP19-0019, 19-ATP19-0020, 19-ATP19-0167, and NSF grants AST-1814053, AST-1814259, AST-1909831, AST-2007355 and AST-2107724.
The simulations were performed on the ``Engaging'' cluster supported by the MIT Kavli Institute and the ``Raven'' HPC system supported by the Max Planck Computing and Data Facility (MPCDF).
The figures in this work were produced using the matplotlib graphics environment \citep{matplotlib2007}, heavily relying on the numpy package \citep{numpy2011}.

\section*{Data Availability}
The data underlying this paper will be shared upon reasonable request to the corresponding author.



\bibliographystyle{mnras}
\bibliography{main.bbl} 




\appendix

\section{Comparison of communication schemes on a typical University cluster}
\label{app:scalingEngaging}
\begin{figure}
    \centering
    \includegraphics[width=1\linewidth]{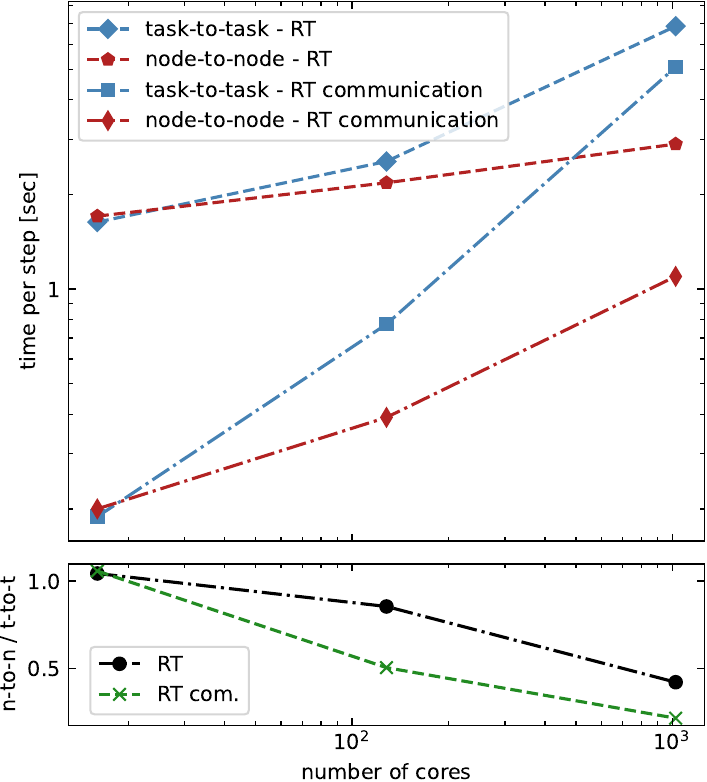}
    \caption{\textbf{Task-to-Task vs Node-to-Node communication, weak scaling:} The average run time per time step as a function of cores for simulating an expanding \HII region on MIT's ``Engaging'' cluster.
    We increase the number of cells with an increasing number of cores to perform a weak scaling test and run the simulation with the new (n-to-n) and the old (t-to-t) communication schemes.
    In the top panel, we show the run time spent in the RT solver and the costs of the MPI communication in the RT solver. Additionally, we show the relative run time difference for both measurements in the bottom panel. For more details, see App.~\ref{app:scalingEngaging}.}
    \label{fig:weak_scaling_hii_engaging}
\end{figure}

As discussed in Section~\ref{subsec:expandingHIIRegionCommunication}, the scalability of a parallelized program depends both on the software and hardware employed to interconnect the computing nodes.
To further investigate this, we replicated the expanding \HII region test outlined in Section~\ref{subsec:expandingHIIRegionCommunication} on the ``Engaging'' cluster at MIT. 
This cluster features dual AMD EPYC 7542 32-core processors per node, equating to 64 cores per node. These nodes are interconnected via a 26Gbit/s InfiniBand, notably slower than the 100 Gbit/s on ``Raven''.
For both weak and strong scaling analyses, we simulated several \HII regions with $2 \times 44^3$ cells. Due to the cluster's limitations, with a maximum of 16 nodes per simulation, we used one \HII region per 16 cores for weak scaling and 8 \HII regions for strong scaling. As shown in \cref{fig:weak_scaling_hii_engaging}, both communication schemes exhibit similar performance when the simulation is confined to a single node. However, as the number of nodes increases, the node-to-node (n-to-n) scheme demonstrates notably better scalability, resulting in more than a twofold increase in full RT performance for 1024 cores compared to the older task-to-task (t-to-t) scheme.

\begin{figure}
    \centering
    \includegraphics[width=1\linewidth]{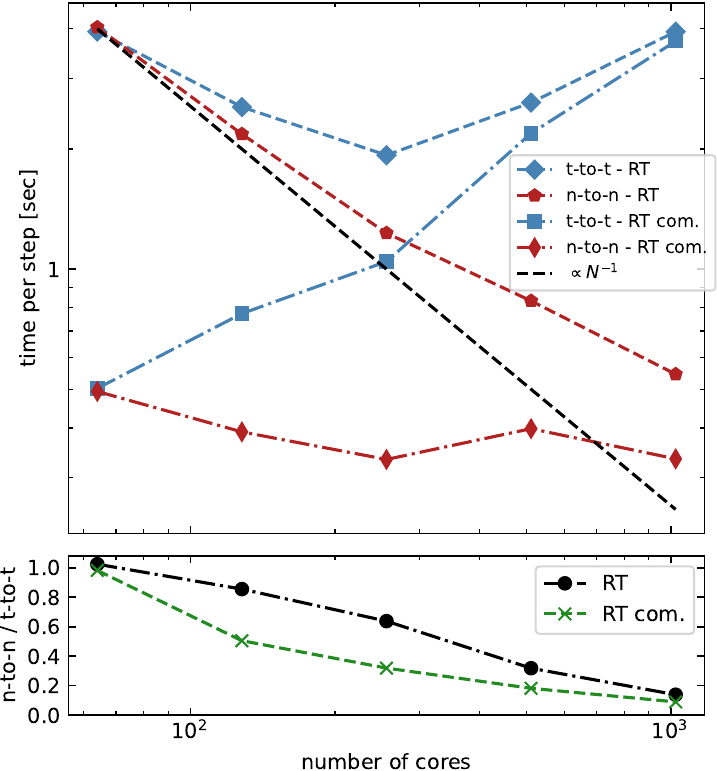}
    \caption{\textbf{Task-to-Task vs Node-to-Node communication, strong scaling:} The same as \cref{fig:weak_scaling_hii_engaging} but for strong scaling, which means we keep the number of cells in the simulation constant $8\times 2 \times 44^3$.}
    \label{fig:strong_scaling_hii_engaging}
\end{figure}

This difference becomes even more extreme for the strong scaling test, as illustrated in \cref{fig:strong_scaling_hii_engaging}.
With the n-to-n scheme, total communication time decreases for a few nodes as the theoretical maximum bandwidth increases. Conversely, with the old t-to-t scheme, communication costs consistently increase with an increasing number of computing nodes, invariably dominating the computing costs for $\geq 512$ cores. Notably, the disparities between the two schemes on this cluster are substantially larger than those identified in Section~\ref{subsec:expandingHIIRegionCommunication}. Thus, the new scheme proves beneficial not only for extensive simulations but also for smaller ones, especially when employing a less-optimized network.

\bsp	
\label{lastpage}
\end{document}